\documentclass [12pt,a4paper] {article}
\usepackage[backend=biber]{biblatex}
\usepackage{amssymb,amsmath} 	% math package
\author {a.author}
\usepackage{graphicx}
\usepackage{subcaption}
\usepackage{setspace}
\usepackage{array}
\usepackage{epsf, color, verbatim}
\usepackage{amssymb,amsmath}
\usepackage{soul}
\usepackage{color}
\usepackage{epsf, color, verbatim}
\usepackage{amssymb,amsmath}
\usepackage{soul}
\usepackage{color}
\usepackage[usenames,dvipsnames]{xcolor}
\usepackage{caption} 
\usepackage{booktabs}
\usepackage[ruled,vlined]{algorithm2e}
\usepackage{tabulary}
\usepackage{biblatex}

\newcommand{\bfmath}[1]{\mbox{\boldmath$#1$\unboldmath}}
\newtheorem{assumption}{Assumption}
\newenvironment{proof}[1][Proof]{\noindent\textbf{#1.} }{\ \rule{0.5em}{0.5em}}
\newtheorem{theorem}{Theorem}%[section]
\newtheorem{lemma}{Lemma}%[section] 
\title{
\textbf{Phase I Analysis of High-Dimensional Processes in the Presence of Outliers}}
\small{\author{Mohsen Ebadi \\
 \makeatother
  {\small \it  University of Waterloo, Waterloo, ON N2L 3G1, Canada} \\
 Shoja'eddin Chenouri 
  \\{\small \it University of Waterloo, Waterloo, ON N2L 3G1, Canada} \\
Stefan H. Steiner\\  {\small \it  University of Waterloo, Waterloo, ON N2L 3G1, Canada}
}
\date{}
}
\begin {document}

\maketitle
\begin {abstract}
One of the significant challenges in monitoring the quality of products today is the high dimensionality of quality characteristics. In this paper, we address Phase I analysis of high-dimensional processes with individual observations when the available number of samples collected over time is limited. Using a new charting statistic, we propose a robust procedure for parameter estimation in Phase I. This robust procedure is efficient in parameter estimation in the presence of outliers or contamination in the data. A consistent estimator is proposed for parameter estimation and a finite sample correction coefficient is derived and evaluated through simulation. We assess the statistical performance of the proposed method in Phase I in terms of the probability of signal criterion. This assessment is carried out in the absence and presence of outliers. We show that, in both phases, the proposed control chart scheme effectively detects various kinds of shifts in the process mean. Besides, we present two real-world examples to illustrate the applicability of our proposed method.
\end {abstract}
\section{Introduction}
In industrial practice, control charts are widely used for monitoring quality characteristics. The quality of a product in many industries is often related to several correlated characteristics, and their combined effect describes product quality. Multivariate control charts have been used for such situations. Because a collection of separate univariate control charts ignores the correlation between the quality characteristics, they do not provide the whole picture of the quality of products. The $T^{2}$ statistic proposed by Hotelling (1947) is a well-known tool for monitoring the mean vector of multivariate processes, and most of the early papers on this topic concentrated on the $T^{2}$ statistic due to its ease of computation. However, for situations where the number of variables is greater than the number of observations, the sample covariance matrix used in the $T^{2}$ statistic is singular. In this case, the methods based on $T^{2}$ perform poorly and become unreliable. Woodall and Montgomery (2014) pointed out that monitoring high-dimensional data, such as those coming from sensors, social networks, health multistream systems, and cloud computing applications, is a complex and important area of research. With advances in computing and modern data-acquisition equipment, high dimensional data are common in many environments, which brings severe challenges to the applicability of the traditional multivariate monitoring methods. Although the multivariate control charting methods are abundant (see and Bersimis et al. (2007) and Ebadi et al. (2021a) for literature review), for monitoring changes in the mean vector of high dimensional multivariate processes the literature is sparse. Moreover, the few existing work mainly focused on the problem of Phase II monitoring of high dimensional processes which will be explained as follows. Variable selection (VS)-based control charts were recently introduced for monitoring multivariate data based on the reasonable assumption that assignable causes usually affect only a small portion of the monitored quality characteristics and not all variables simultaneously.
Wang and Jiang (2009) proposed using a forward selection algorithm (FVS) to identify the subset of the shifted variables, combined with a Shewhart-type control chart, referred to as the VS-MSPC chart. Jiang et al. (2012) suggested using a multivariate EWMA (MEWMA) control chart instead of a Shewhart-type control chart to accumulate recent observations and make an integrated procedure VS-MEWMA chart, more sensitive to the small shifts. However, Abdella et al. (2016) explained that variable selection procedure malfunctions might lead to the poor performance of the VS-MEWMA in detecting small process changes and proposed a VS-MCUSUM control chart to improve the detection of small changes in the mean vector. Zou and Qiu (2009) used the least absolute shrinkage and selection operator (LASSO) as a variable selection method and applied the EWMA control chart to develop a LASSO-based EWMA (LEWMA) chart for monitoring the process mean. The LASSO-based schemes assume that the number of potential out-of-control variables is not fixed a priori, and any subset of the monitored variables can potentially shift. Capizzi and Masarotto (2015) compared the performances of three VS-based control charts, including the least angle regression (LAR) control chart proposed by Capizzi and Masarotto (2011), LEWMA, and VS-MEWMA charts. 

Generally, monitoring a multivariate process contains two phases, Phase I and Phase II.  According to Woodall (2000), the primary concern of Phase I is to analyze historical data to understand the underlying variation and determine the stability of the process. Having removed those samples associated with any assignable causes in Phase I, we can estimate the in-control values of the process parameters to use them in designing control charts for Phase II monitoring of the process. On the other hand, the primary purpose of Phase II is to quickly detect shifts in the process parameters from the in-control (IC) parameter values estimated in the Phase I step. A significant challenge for monitoring high-dimensional processes is that only a small reference dataset is often available. At the same time, the dimension $p$ of the Phase I data is greater than the Phase I sample size $m$. Hence, the Phase I analysis can not be performed as usual, and historical observations are insufficient to accurately estimate all parameters of the in-control distribution. Besides, outliers may exist in Phase I data, which makes this analysis even harder. There is a need to develop more robust estimators of the process parameters in the high-dimensional setting. We illustrate the issue using a dataset of vertical density profiles (VDP) of engineered wooden boards. This dataset has been studied by several authors such as Walker and Wright (2002), Woodall et al. (2004), Williams et al. (2007) and Wang and Jiang (2009). 

In many practical situations, the covariance matrix can not be meaningfully estimated from the original data due to the ``curse of dimensionality”. In this paper, we apply a robust method to estimate the parameters and propose a new control chart in Phase I.  In our new chart, rather than estimating all elements of the covariance matrix, we only estimate it's diagonal elements. We show our approach is very effective in Phase I analysis of the process mean, especially when the sample size is small compared to the number of variables.
The rest of this paper is organized as follows. In Section 2, we develop our proposed charts based on the diagonal elements of the sample covariance matrix. We use a robust estimation procedure to construct the proposed chart in this section and present a finite sample correction coefficient in the Phase I analysis. Using Monte Carlo simulations, the performance of the proposed chart is evaluated in terms of probability of signal in Section 3. In Section 4, two illustrative examples are investigated to demonstrate the performance of the proposed methodology. Our conclusions are summarized in Section 5.

\section{Proposed Method}
In this section, we begin by proposing a control chart for high-dimensional processes and then present an estimation approach for the underlying parameters of the in-control processes in Phase I. We further propose a finite sample correction coefficient to be used for the Phase I analysis. 
%%%%%%%%%%%%%%%%%%%%%%%%%%%%%%%%%
\subsection{Charting statistic and robust parameter estimation}
Suppose the multivariate process with $p$ quality characteristics $\mathbf{X}=(X_1,\,X_2,\,\dots,\,X_p)'$ and there are $m$ independent and identically distributed (i.i.d) historical (reference) observations $\mathbf{X}_{1},\,\mathbf{X}_{2},\,\dots,\,\mathbf{X}_{m} $ in Phase I. The process follows a multivariate normal distribution with mean vector $\bfmath{\mu}$ and covariance matrix $\bfmath{\Sigma}$ under an in-control situation. In practice, the available number of Phase I samples is usually limited, i.e. $p>m$, and the standard sample covariance matrix is singular such that the typical approaches based on $T^{2}$ statistic becomes impractical. Assume the sampling epoch $i$ and denote the individual observation of the $j$th quality characteristic variable by $X_{i\,j},$ where $i=1,\,\dots\,m$, and $j=1,\,\dots,\,p$. Let $\bfmath{\sigma}=({\sigma}_{_{1\,1}},\,\dots,\,{\sigma}_{_{p\,p}})'$ denote the vector of in-control variances of the $p$ variables obtained from the diagonal elements of $\bfmath{\Sigma}$. If we define $\bfmath{D}=\rm{diag}({\sigma}_{_{1\,1}},\,\dots,\,{\sigma}_{_{p\,p}})$, then the corresponding modified Mahalanobis distance is:
\begin{equation}\label{e2}
M^{2}_{i}=M^{2}_{i}(\bfmath{\mu},\mathbf{D})=({\mathbf{X}_{i}}-\bfmath{\mu})'\mathbf{D^{-1}}({\mathbf{X}_{i}}-\bfmath{\mu})=\sum_{j=1}^{p}\frac{(X_{ij}-{\mu}_{_j})^{2}}{\sigma_{_{j\,j}}}\,,
\end{equation}
where ${\mu}_{_{j}}$ denotes the $j$th element of the vector $\bfmath{\mu}$. Similar to Ro et al. (2015), we call $M^{2}_{i}$ modified Mahalanobis distance, since the covariance matrix $\bfmath{\Sigma}$ has been replaced with the diagonal matrix $\mathbf{D}$
in its formula.
A Phase I control chart, however, can be obtained by replacing $\bfmath{\mu}$ and $\bfmath{\Sigma}$ with the sample mean and covariance matrix, respectively (Bersimis et al. 2007). Let $\lambda_{1},\,\lambda_{2}, ..., \,\lambda_{p}$ denote the eigenvalues of the in-control correlation matrix  $\bfmath{\rho}= \mathbf{D}^{-\frac{1}{2}}\,\bfmath{\Sigma}\,\mathbf{D}^{-\frac{1}{2}}$, then the mean and variance of $M^{2}_{i}$ are given by (see Ebadi et al. 2021b for the proof)
\begin{equation}\label{e3}
{\rm E}(M^{2}_{i})=p, \, \, {\rm Var}\left(M^{2}_{i}\right)=2\,\textrm{tr}(\bfmath{\rho}^2),
\end{equation}
where $\textrm{tr}(\bfmath{A})$ represents the trace of matrix $\bfmath{A}$. Hence, the following test statistic can be defined:
\begin{equation}\label{e5}
U_{i}=\frac{M^{2}_{i}(\bfmath{\mu},\mathbf{D})-p}{\sqrt{2\,{\rm tr}(\bfmath{\rho}^2)}}\,.
\end{equation}
It can be shown that for any given $i=1,\,\dots,\,m$ the statistic $U_{i}$ has an asymptotic $N(0,1)$ distribution as $p\rightarrow\infty$ under the following assumptions: 

\begin{assumption}\label{A1} 
For $i=1,\,\dots,\,6\,$, we assume that  $0<\lim\limits_{p\rightarrow{\infty}}\,p^{-1}\,{\rm tr}\left(\bfmath{\rho}^i\right)<\infty\,$.
\end{assumption}

\begin{assumption}\label{A2}
The eigenvalues $\lambda_i$ of the correlation matrix $\bfmath{\rho}$ satisfy $\lim\limits_{p\rightarrow{\infty}} \underset{1\leq i\leq p}{\max} \,\lambda_i\, p^{-1/2}=0\,$.
\end{assumption}

\begin{assumption}\label{A3}
The dimension $p$ grows with sample size $m$ at a rate of $p=O(m^{1/\zeta})$ with $1/2<\zeta\leq{1}\,$.

\end{assumption}

\begin{assumption}\label{A4}
For some $0 <\gamma <\zeta/2$, $\lim\limits_{p\rightarrow{\infty}} \underset{1\leq i\leq p}{\max}\,\lambda_i\,p^{-\gamma}<\infty\,$.
\end{assumption}
A discussion on these assumptions is provided in Ebadi et al. 2021b. To provide better accuracy in tails when small values of $\alpha$ such as 0.005 are used in constructing control charts, one can modify the statistic by using the first-order expansion of the Cornish-Fisher
\begin{align}\label{CFexp1}
\omega_{\alpha,\,p}\approx z_{\alpha}+\frac{4\,{\rm tr}(\bfmath{\rho}^{3})\,(z_{\alpha}^2-1)}{3\left[2\,{\rm tr}(\bfmath{\rho}^{2})\right]^\frac{3}{2}}\,,
\end{align}
as it suffices to achieve good results (Ebadi et al. 2021b).

%So, if we define the new variable $Z_i$ at a given  significance level $\alpha$ by
%\begin{equation}\label{e8}
%Z_i=U_i-\frac{(4\textrm{tr}\bfmath{\rho}^{3})(z_{\alpha}^2-1)}{3(2\textrm{tr}\bfmath{\rho}^{2})^\frac{3}{2}}
%\end{equation}
%then, for large dimension $p$ $\Pr\left[-z_{\alpha/2}\leq {Z_i}\leq{z_{\alpha/2}}\right]$ more accurately approximates $=1-\alpha$ in comparison to $\Pr\left[-z_{\alpha/2}\leq {U_i}\leq{z_{\alpha/2}}\right]$. 
Hence, the modified charting statistic triggers an out-of-control alarm whenever 
\begin{equation}\label{chart}
Z_{i}=U_{i}-\frac{4\,\textrm{tr}(\bfmath{\rho}^{3})\,(z_{\alpha}^2-1)}{3\left[2\,{\rm tr}(\bfmath{\rho}^{2})\right]^\frac{3}{2}}>z_{\alpha},
\end{equation}

%Assume that proper and robust estimates of mean vector $\mathbf{\mu}$, diagonal matrix $D$, $\textrm{tr}\bfmath{\rho}^{2}$, and $\textrm{tr}\bfmath{\rho}^{3}$ are obtained based on $m$ observations in Phase I analysis (we will discuss this in more details in next subsection) such that they are very close to the actual parameters and new individual observations $X_{m+1},X_{m+2},...,$ are to be monitored in Phase II by using control limit (7). 

It is worth mentioning here that recently, Martinez et al (2020) used a Bisection algorithm to numerically find the empirical control limits of RMDP based on simulation. A similar approach was used in Chenouri et al. (2009) for RMCD method, where they used the empirical distribution function of
$T^2_{RMCD}$ and obtained Monte Carlo estimates of the
99\%, and 99.9\% quantiles. However, such simulation-based methods should be performed for any (different) values of $m$ and $p$ to find the corresponding thresholds. When the process is in-control, a new observation follows $N_{p}(\bfmath{\mu},\bfmath{\Sigma})$, while in the out-of-control situation the observations follow $N_{p}(\bfmath{\mu}_1,\bfmath{\Sigma})$.
Consequently, assuming $\bfmath{\mu}_1-\bfmath{\mu}=\bfmath{\delta}$, the asymptotic Type II error probability of the proposed control chart as $p\rightarrow\infty$ can be derived as (see Ebadi et al. 2021b for more details and the proof.)

\begin{align*}
 \lim\limits_{p\rightarrow \infty}\Pr(U_i\le z_\alpha\,\mid \bfmath{\mu}_1)&=
  \lim\limits_{p\rightarrow\infty}
 \Pr\left(\frac{M^{2}_{i}(\bfmath{\mu}_{1},\mathbf{D})+\bfmath{\delta}^{\prime}\mathbf{D}^{-1}\bfmath{\delta}-p}{\sqrt{2\,{\rm tr}(\bfmath{\rho}^{2})}}\le z_\alpha\,\mid \bfmath{\mu}_1\right) \\ 
 &= \lim\limits_{p\rightarrow\infty}
 \Phi\left(z_\alpha
-\frac{{\bfmath{\delta}}^{\prime}\mathbf{D}^{-1}\bfmath{\delta}}{\sqrt{2\,{\rm tr}(\bfmath{\rho}^{2})}}\right) \,. 
\end{align*}

As mentioned, proper estimates of the parameters, $\bfmath{\mu}$,  $\mathbf{D}$, ${\rm tr}(\bfmath{\rho}^{2})$, and ${\rm tr}(\bfmath{\rho}^{3})$ in Phase I are needed. We now discuss the methodology for the robust estimation of the parameters of the proposed control chart.  It is known that, under the normality assumption of the Phase I data, the maximum likelihood estimates of these parameters are optimal. However, it is also well known that these estimators, sometimes referred to as the classical estimators, are sensitive to outlying observations in the Phase I dataset. This section presents an algorithm for robust estimation of the Phase I parameters. We first introduce consistent estimators for ${\rm tr}(\bfmath{\rho}^{2})$ and ${\rm tr}(\bfmath{\rho}^{3})$, and then briefly explain the algorithm. Recall that there are $m$ i.i.d. historical observations $\mathbf{X}_{1},\,\mathbf{X}_{2},\,\dots,\,\mathbf{X}_{m}$ collected for Phase I analysis. The sample correlation matrix $\bfmath{R}$ in Phase I,  based on $m$ observations, can be represented by
\begin{equation}\label{e6} 
 \bfmath{R}= \bfmath{D}_{_S}^{-\frac{1}{2}}\,\bfmath{S}\,\bfmath{D}_{_S}^{-\frac{1}{2}}\,,
\end{equation}
where $\bfmath{S}$ is the sample covariance matrix and $\bfmath{D}_{_S}$ denotes the diagonal matrix of the sample variances in $\bfmath{S}$. In practice, $\bfmath{\rho}$ is unknown. To use \eqref{e5} and \eqref{chart}, one needs to estimate ${\rm tr}(\bfmath{\rho}^{2})$ and ${\rm tr}(\bfmath{\rho}^{3})$, in addition to $\bfmath{\mu}$ and $\mathbf{D}$. From classical multivariate statistics, we know that for a fixed $p$, ${\rm tr}(\bfmath{R}^{2})$ and ${\rm tr}(\bfmath{R}^{3})$ are consistent estimators of ${\rm tr}(\bfmath{\rho}^{2})$ and ${\rm tr}(\bfmath{\rho}^{3})$, respectively, as $m\rightarrow \infty$. However, this is not the case when both $m$ and $p$ grow unbounded. Motivated by the results in Bai and Saranadasa (1996) and Srivastava (2005), the paper Srivastava and Du (2008) proposed $m^2\,\left[(m-1)\,(m+2)\,p\right]^{-1}\left[{\rm tr}(\bfmath{R}^{2})-p^2/m\right]$ as the uniformly minimum variance unbiased estimator (UMVUE) of $p^{-1}{\rm tr}(\bfmath{\rho}^{2})$. Furthermore, since $m^2/(m-1)\,(m+2)\approx 1$ when $m\rightarrow \infty$, Srivastava and Du (2008) suggested to simply use $p^{-1}\left[{\rm tr}(\bfmath{R}^{2})-p^2/m\right]$ and additionally established its consistency under Assumption 1 and 3, in the sense that  
$$\frac{1}{p}\left[{\rm tr}(\bfmath{R}^{2})-\frac{p^2}{m}\right]-\frac{1}{p}{\rm tr}(\bfmath{\rho}^{2})\rightarrow 0 \quad \text{as } m, \,p \rightarrow\infty\,.$$

Srivastava and Du (2008) were concerned with a hypothesis testing problem involving the mean vector of a multivariate normal distribution. They used a test statistic similar to $U_i$ in \eqref{e5} when in its denominator, the parameter  $2\,{\rm tr}(\bfmath{\rho}^2)$ is replaced by an estimator in the form of 
$2\,c^\ast_{{p,\,m}}\left[{\rm tr}(\bfmath{R}^{2})-p^2/m\right]$, where $c^\ast_{{p,\,m}}\overset{p}\longrightarrow 1$ as $m,\,p\rightarrow \infty$. Srivastava and Du (2008) also argued that $c^\ast_{{p,\,m}}=1+p^{-3/2}\,{\rm tr}(\bfmath{R}^{2})$ leads to a faster convergence to
normality. As an alternative, in this paper for process monitoring, we propose an alternative coefficient
based on the distributional properties of estimators. In the next subsection and Appendix B, via a simulation study and careful numerical evaluations, we propose using
\begin{equation}\label{e11}
c_{_{p,\,m}}=1+\frac{2\,p}{m\sqrt{{\rm tr}(\bfmath{R}^{2})-\frac{p^2}{m}}}
\end{equation} 
instead. It is easy to show that under Assumptions 1 and 3, $c_{_{p,\,m}}\overset{p}\longrightarrow 1$ as $m,\,p\rightarrow \infty$.  

Recall that we introduced a Cornish-Fisher expansion for the distributional quantiles of $U_i$. This result leads to the threshold \eqref{chart}, which is being used to  chart statistic $Z_i$. To calculate the threshold \eqref{chart} in practice, we must provide an estimate of ${\rm tr}(\bfmath{\rho}^{3})$ based on the Phase I data. The following Theorem provides a consistent estimator of ${\rm tr}(\bfmath{\rho}^{3})$.

\begin{theorem}\label{R3} Under Assumptions 1 and 3, $p^{-1}\left[\,{\rm tr}(\mathbf{R}^3)-3\,m^{-1}\,p\,{\rm tr}(\mathbf{R}^2)+2\,m^{-2}\,p^3\,\right]$ converges to $\lim\limits_{p\rightarrow{\infty}}p^{-1}{\rm tr}(\bfmath{\rho}^{3})$ in probability as $(m,p)\rightarrow\infty$. So, a consistent estimator of ${\rm tr}(\bfmath{\rho}^{3})$ is given by
\begin{equation}\label{e8}
{\rm tr}(\mathbf{R}^3)-\frac{3p}{m}\,{\rm tr}(\mathbf{R}^2)+\frac{2\,p^3}{m^2}
\end{equation}
\end{theorem}
A proof for Theorem \ref{R3} is provided in Appendix A.

Under the Assumptions 1, 3 and 4, and outlier free data, Ro et al. (2015) showed that when the classical estimators denoted by $\widehat{\bfmath{\mu}}$ and $\widehat{\mathbf{D}}$ are used, the following result holds
\begin{equation}\label{Mhat-M}
\max_{{1\leq i\leq m}}\left|\frac{M^{2}_{i}(\widehat{\bfmath{\mu}},\widehat{\mathbf{D}})-p}{\sqrt{2\,{\rm tr}(\bfmath{\rho}^{2})}}-\frac{M^{2}_{i}(\bfmath{\mu},\mathbf{D})-p}{\sqrt{2\,{\rm tr}(\bfmath{\rho}^{2})}}\right|=o_{p}(1),\quad \quad   \textrm{as} \quad    m,\,p\rightarrow\infty
\end{equation}
Although the result \eqref{Mhat-M} was proved for outlier-free data, it is also valid when $\widehat{\bfmath{\mu}}$ and $\widehat{\mathbf{D}}$ are replaced by consistent robust estimators when data are contaminated with outliers. In order to obtain robust estimators of parameters in the hypothesis testing problem of location, Ro et al. (2015) introduced re-weighted minimum diagonal product (RMDP) estimators by modifying the re-weighted minimum covariance determinant (RMCD) algorithm of Rousseeuw and Van Driessen (1999). The current paper proposes a robust estimation procedure for parameter estimation in Phase I, similar to Ro et al. (2015). However, due to the use of the Cornish-Fisher expansion, we additionally must estimate  ${\rm tr}(\bfmath{\rho}^{3})$  and modify all computational steps of the algorithm accordingly to accommodate for this change. This modification makes our algorithm quite different from that of Ro et al. (2015). In what follows, we provide a summary of our proposed re-weighting algorithm. 

Motivated by the RMCD estimators of Rousseeuw and Van Driessen (1999), we start by searching for a subset of size $h$ in Phase I data with the smallest product of the diagonal elements of the sample covariance matrix. The integer $h=\lfloor m\,(1-\gamma)\rfloor$ is the sample size to achieve $100\,\gamma\%$ breakdown value for the output estimators, where $0\le \gamma\le 0.5$. See Rousseeuw and Van Driessen (1999). More formally, let $|H|$ denote the size or cardinality of the set $H$ and $\mathcal{H}=\left\lbrace H \subset \lbrace 1,\,\dots,\,m\rbrace:\,|H|=h\right\rbrace$. For any $H \in \mathcal{H}$ and the sub-sample $\lbrace{\mathbf{X}_{i} :i \in {H}}\rbrace$, let $\widehat{\bfmath{\mu}}(H)$ and $\widehat{\bfmath{\Sigma}}(H)$ denote the sub-sample mean and covariance matrix, respectively. The minimum diagonal product (MDP) estimator of the mean vector $\bfmath{\mu}$ is defined as
 \begin{equation}\label{MDPmean}
\widehat{\bfmath{\mu}}_{_{\rm MDP}}=\widehat{\bfmath{\mu}}(H_{_{\rm MDP}})\,,\quad \text{ where } \quad H_{_{\rm MDP}}=\underset{H \in \mathcal{H}}{\operatorname{argmin}}\,\, {\rm det}\left({\rm \bf diag}\left(\widehat{\bfmath{\Sigma}}(H)\right)\right)\,,
 \end{equation}
Notice that ${\rm\bf diag}\left(\widehat{\bfmath{\Sigma}}(H)\right)$ denotes the matrix of diagonal elements of $\widehat{\bfmath{\Sigma}}(H)$. The MDP estimator of the diagonal matrix $\mathbf{D}$ is given by
  \begin{equation}\label{e8}
\widehat{\mathbf{D}}_{_{\rm MDP}}=a_{\gamma,\,p}^m\,{\rm \bf diag}\left(\widehat{\bfmath{\Sigma}}(H_{_{\rm MDP}})\right)\,, 
 \end{equation}
where $a_{\gamma,\,p}^m$ is a scaling factor depending on $\gamma$, $p$, and $m$ to ensure consistency and finite sample accuracy of $\widehat{\mathbf{D}}_{_{\rm MDP}}$ for multivariate normal data. See Croux and Haesbroeck (1999) and Pison et al. (2002). Moreover, we propose the the following MDP estimators of ${\rm tr}(\bfmath{\rho}^2)$ and ${\rm tr}(\bfmath{\rho}^3)$ 
 \begin{align}
 \widehat{{\rm tr}\,(\bfmath{\rho}^2)}_{_{\rm MDP}}&={\rm tr}\,(\mathbf{R}_{_{\rm MDP}}^2)-\frac{p^2}{h}\,,\label{RobTr2}\\
 \widehat{{\rm tr}\,(\bfmath{\rho}^3)}_{_{\rm MDP}}&={\rm tr}\,(\mathbf{R}_{_{\rm  MDP}}^3)-\frac{3\,p}{h}\,{\rm tr}\,(\mathbf{R}_{_{\rm MDP}}^2)+\frac{2\,p^3}{h^2}\,,\label{RobTr3}
 \end{align}
where $\mathbf{R}_{_{\rm MDP}}$ is the correlation matrix associated with $\widehat{\bfmath{\Sigma}}(H_{_{\rm MDP}})$.
 Algorithm 1 of Ro et al. (2015), adapted the fast minimum covariance determinant (MCD) algorithm of Rousseeuw and Van Driessen (1999) to obtain the MDP estimates of $\bfmath{\mu}$ and $\bfmath{D}$. In the current paper, we use the MDP estimates of Ro et al. (2015) as initial estimates and provide adjusted or re-weighted MDP estimates incorporating the Cornish-Fisher expansion. Our reweighed algorithm below is motivated by Rousseeuw and Van Driessen (1999) and Ro et al. (2015).
\\

\noindent 
{\bf Algorithm 1: Reweighted MDP with Cornish-Fisher expansion}
 \begin{itemize}
\item[]{\it Step 1.} Initialize $\alpha$ and $h$, for example $\alpha=0.05$ and $h=[m/2]+1$ .   
\item[]{\it Step 2.} Run Algorithm 1 in Ro et al. (2015) and obtain the raw MDP estimates $\widehat{\bfmath{\mu}}_{_{\rm MDP}}$ and $\widehat{\mathbf{D}}_{_{\rm MDP}}$, and then calculate $\widehat{{\rm tr}(\bfmath{\rho}^2)}_{_{\rm MDP}}$ and $\widehat{{\rm tr}\,(\bfmath{\rho}^3)}_{_{\rm MDP}}$ using equations \eqref{RobTr2} and \eqref{RobTr3}, respectively to provide initial raw MDP estimates of ${\rm tr}(\bfmath{\rho}^{2})$, ${\rm tr}(\bfmath{\rho}^{3})$. 
\item[]{\it Step 3.} Calculate the value of $M^{2}_{i}(\widehat{\bfmath{\mu}}_{_{\rm MDP}},\widehat{\bfmath{D}}_{_{\rm MDP}})$ for $i=1,\dots,m$. From equations \eqref{e5} and \eqref{chart}, the observation $\mathbf{X}_i$ is identified as an outlier if
\begin{equation}\label{step3}
\frac{M_i^2(\widehat{\bfmath{\mu}}_{_{\rm MDP}},\widehat{\mathbf{D}}_{_{\rm MDP}})-p}{\left[{2\,c^{^{MDP}}_{p,m}\,\widehat{{\rm tr}\,(\bfmath{\rho}^2)}_{_{\rm MDP}}}\right]^{1/2}}-\frac{4\,\widehat{{\rm tr}\,(\bfmath{\rho}^3)}_{_{\rm MDP}}\,(z_{_{\alpha/2}}^2-1)}{3\,\left[{2\,\widehat{{\rm tr}\,(\bfmath{\rho}^2)}_{_{\rm MDP}}}\right]^\frac{3}{2}}>z_{_{\alpha/2}},
  \end{equation}
where 
\begin{equation}\label{cMDP}
c^{^{\rm MDP}}_{p,m}=1+\frac{2\,p}{m\sqrt{{\widehat{{\rm tr}(\bfmath{\rho}^2)}_{_{\rm MDP}}}}}\,.
\end{equation}

Note that the left side of \eqref{step3} is obtained by replacing $\widehat{\bfmath{\mu}}_{_{\rm MDP}}$,  $\widehat{\mathbf{D}}_{_{\rm MDP}}$, $c^{^{\rm MDP}}_{p,m}$, and $\widehat{{\rm tr}(\bfmath{\rho}^2)}_{_{\rm MDP}}$ in \eqref{e5} and deriving the corresponding Cornish-Fisher terms. We used significance level $\alpha/2$ at this step similar to step 2 of  RMDP algorithm in Ro et al. (2015). Also note that \eqref{cMDP} is obtained by replacing $\widehat{{\rm tr}(\bfmath{\rho}^2)}_{_{\rm MDP}}$ in \eqref{e11}, since it is the proper estimate of ${\rm tr}(\bfmath{\rho}^2)$ based on the observations at this step.   
\item[]{\it Step 4.} Assign a weight to each observation according to the threshold in \eqref{step3} by $\omega_i=0$ if the $i$-th observation is identified as an outlier and $\omega_i=1$,  otherwise. Let $\widetilde{\bfmath{\mu}}$, $\widetilde{\mathbf{D}}$, and $\widetilde{\bfmath{\rho}}$ represent the sample mean, diagonal matrix of variances and the correlation matrix, respectively, associated with the sample covariance matrix $\widetilde{\bfmath{\Sigma}}$ computed from all observations $\mathbf{X}_{i}$ for which $\omega_i=1$.
\item[]{\it Step 5.} Denote the re-weighted estimates of $\mathbf{D}$, ${\rm tr}(\bfmath{\rho}^2)$, and ${\rm tr}(\bfmath{\rho}^3)$ by $\widehat{\mathbf{D}}_{_{\rm RMDP}}$, $\widehat{{\rm tr}\,(\bfmath{\rho}^2)}_{_{\rm RMDP}}$, and $\widehat{{\rm tr}\,(\bfmath{\rho}^3)}_{_{\rm RMDP}}$, respectively and update the traces through 
\begin{align}
\widehat{{\rm tr}\,(\bfmath{\rho}^2)}_{_{\rm RMDP}}&={\rm tr}\,(\widetilde{\bfmath{\rho}}^2)-\frac{p^2}{\sum_{i=1}^{m}\omega_i}\,\,, \label{RMPD2}\\
\widehat{{\rm tr}\,(\bfmath{\rho}^3)}_{_{\rm RMDP}}&={\rm tr}\,(\widetilde{\bfmath{\rho}}^3)-\frac{3\,p}{\sum_{i=1}^{m}\omega_i}\cdot{\rm tr}\,(\widetilde{\bfmath{\rho}}^2)+\frac{2\,p^3}{\left(\sum_{i=1}^{m}\omega_i\right)^2}\,\,.\label{RMPD3}
\end{align} 
Since an updated $\widehat{\mathbf{D}}_{_{\rm RMDP}}$ is not available until the end of Step 6, for the $i$-th observation with $\omega_i=1$, using the Proposition 2 in Ro et al. (2015), compute the refined distance 
\begin{equation}\label{e8}
M^{2}_{i}(\widetilde{\bfmath{\mu}}, \,\widehat{\mathbf{D}}_{_{\rm RMDP}})\approx\frac{M^{2}_{i}(\widetilde{\bfmath{\mu}},\, \widetilde{\mathbf{D}})}
{1+\phi(z_{_{\alpha/2}})\,p^{-1}(1-\alpha/2)^{-1}\,\sqrt{2\,\widehat{{\rm tr}\,(\bfmath{\rho}^2)}_{_{\rm RMDP}}}}\,,
 \end{equation}
where $\phi$ is the standard normal density function. Then, evaluate the outlyingness of each observation with the rejection region 
 \begin{equation}\label{tresh}
 \frac{M^{2}_{i}(\widetilde{\bfmath{\mu}}, \,\widehat{\mathbf{D}}_{_{\rm RMDP}})-p}{\left[2\,\tilde{c}_{p,m}\,\widehat{{\rm tr}\,(\bfmath{\rho}^2)}_{_{\rm RMDP}}\right]^{1/2}}-\frac{4\,(z_{_{\alpha}}^2-1)\,\widehat{{\rm tr}(\bfmath{\rho}^3)}_{_{\rm RMDP}}}{3\,\left[2\,\widehat{{\rm tr}(\bfmath{\rho}^2)}_{_{\rm RMDP}}\right]^\frac{3}{2}}>z_{\alpha}\,,
 \end{equation} 
  where $\tilde{c}_{p,\,m}$ is obtained from \eqref{cMDP} by substituting  $\widehat{{\rm tr}(\bfmath{\rho}^2)}_{_{\rm MDP}}$ with $\widehat{{\rm tr}(\bfmath{\rho}^2)}_{_{\rm RMDP}}$.
\item[{\it Step 6.}] Update weights of observations based on their outlyingness determined in Step 5. Calculate the process parameters by updating $\widetilde{\bfmath{\mu}}$, $\widetilde{\bfmath{D}}$, $\widehat{{\rm tr}\,(\bfmath{\rho}^2)}_{_{\rm RMDP}}$, and $\widehat{{\rm tr}\,(\bfmath{\rho}^3)}_{_{\rm RMDP}}$ based on the new observations weights.

 \end{itemize}
% \caption{Re-weighted MDP incorporating Cornish-Fisher Expansion}
 %\end{algorithm}
% {\bf Reweighting Algorithm}

\subsection{Finite sample correction}
In this subsection, we propose a finite sample correction factor for the asymptotic distribution of the test statistic $U_i$. This correction will improve the accuracy of the control limit obtained from the asymptotic normality with Cornish-Fisher expansion in Phase I. 

\begin{lemma}\label{lemma3}
For the usual estimators $\widehat{\bfmath{\mu}}$ and $\widehat{\mathbf{D}}$, we have
\begin{equation}\label{e2}
M^{2}_{i}(\widehat{\bfmath{\mu}},\,\widehat{\mathbf{D}})=M^{2}_{i}(\bfmath{\mu},\,\mathbf{D})+O_{p}\left(\frac{p}{\sqrt{m}}\right)\,,
\end{equation}
where the subscript ‘p’ denotes ‘in probability’.
\end{lemma}
\begin{proof}
Let $s_{ii}, i = 1, \dots , p$ be the $i$th diagonal element of the sample covariance matrix $\mathbf{S}$. From Corollary 2.6 of Srivastava (2009), for large $m$, we have ${\rm E}\left[s_{ii}^{-1}\right]=\sigma_{ii}^{-1}+O(m^{-1})$. Thus, $s_{ii}^{-1}=\sigma_{ii}^{-1}+O_{p}(m^{-1})$ and
\begin{align}\label{Dhat}
\widehat{\mathbf{D}}^{-1}=\mathbf{D}^{-1}\left[1+O_{p}\left(m^{-1}\right)\right]\,.
\end{align}
Since
\begin{equation}\label{Mhat}
M^{2}_{i}(\widehat{\bfmath{\mu}},\,\widehat{\mathbf{D}})=M^{2}_{i}(\bfmath{\mu},\,\widehat{\mathbf{D}})+(\widehat{\bfmath{\mu}}-\bfmath{\mu})'\,\widehat{\mathbf{D}}^{-1}(\widehat{\bfmath{\mu}}-\bfmath{\mu})-2(\mathbf{X}_i-\bfmath{\mu})'\,\widehat{\mathbf{D}}^{-1}(\widehat{\bfmath{\mu}}-\bfmath{\mu})\,,
\end{equation}
using \eqref{Dhat}, we obtain
\begin{align*}
&M^{2}_{i}(\bfmath{\mu},\,\widehat{\mathbf{D}})=M^{2}_{i}(\bfmath{\mu},\,\mathbf{D})+O_{p}\left(p\,m^{-1}\right)\\
&(\widehat{\bfmath{\mu}}-\bfmath{\mu})'\,\widehat{\mathbf{D}}^{-1}(\widehat{\bfmath{\mu}}-\bfmath{\mu})=O_{p}\left(p\,m^{-1}\right)\\
&(\mathbf{X}_i-\bfmath{\mu})'\,\widehat{\mathbf{D}}^{-1}(\widehat{\bfmath{\mu}}-\bfmath{\mu})=O_{p}\left(p\,m^{-1/2}\,\right)\,.
\end{align*}
Substituting these in \eqref{Mhat} we conclude that
$M^{2}_{i}(\widehat{\bfmath{\mu}},\,\widehat{\mathbf{D}})=M^{2}_{i}(\bfmath{\mu},\,\mathbf{D})+O_{p}\left(p\,m^{-1/2}\right)$ 
\end{proof}

Recall the definition of $U_i$ in equation \eqref{e5} and define $\widehat{U}_{i}$, when $M^{2}_{i}(\bfmath{\mu},\,\mathbf{D})$ is substituted by $M^{2}_{i}(\widehat{\bfmath{\mu}},\,\widehat{\mathbf{D}})$ in \eqref{e5}. From Lemma \ref{lemma3}, we can write 
\begin{equation}\label{uhat}
\widehat{U}_{i}=U_{i}\left[1+\frac{O_{p}\left(p\,m^{-1/2}\right)}{\sqrt{{\rm tr}(\bfmath{\rho}^2)}}\right]
\end{equation}
If we replace ${\rm tr}(\bfmath{\rho}^2)$ by its estimate introduced in Section 2.2. we can write $\widehat{U}_{i}=U_{i}\,c_{p,m}$. Our simulation study reported in Appendix B suggests a finite sample correction factor having the form given by Eq.\eqref{e11}.
It is worth mentioning here that although the Cornish-Fisher expansion was based on the normality assumption, the result of Lemma \ref{lemma3} and thus the derivation of the finite sample correction factor does not heavily rely on this assumption. Moreover, Srivastava (2009) showed that the one-sample version of test statistic $U_i$ could also be applied for non-normal data under certain assumptions. So, a natural suggestion for the case of non-normal processes is to use the statistic $U_i$ together with the finite sample correction defined in\eqref{cpm}.
 A future research direction is on improving the performance of the proposed charting statistic for various distributional assumptions. In the next section, we examine the performance of our proposed methods through simulation. 
%%%%%%%%%%%%%%%%%%%%%

\section{Performance evaluation}

In this section, we evaluate the proposed methods through a simulation study both in the absence and presence of contaminated data. We use R (R Development Core Team, 2019) for all the computation. We consider two scenarios, where in both scenarios the common mean vector is $\mathbf{0}_{p}$, while their covariance matrices are $\mathbf{I}_p$, and $\sigma_{ij}=(0.5)^{|i-j|}\quad {\rm for}\quad i,j=1,\dots,p$, respectively which are related to independent and autoregressive correlation structures. We use these two scenarios in this section to evaluate our proposed methods. All the simulation runs are performed based on 10,000 replications unless indicated otherwise.

\begin{figure}[h!]
  \centering
  \begin{subfigure}[b]{0.45\linewidth}
    \includegraphics[width=\linewidth,height=2.5in]{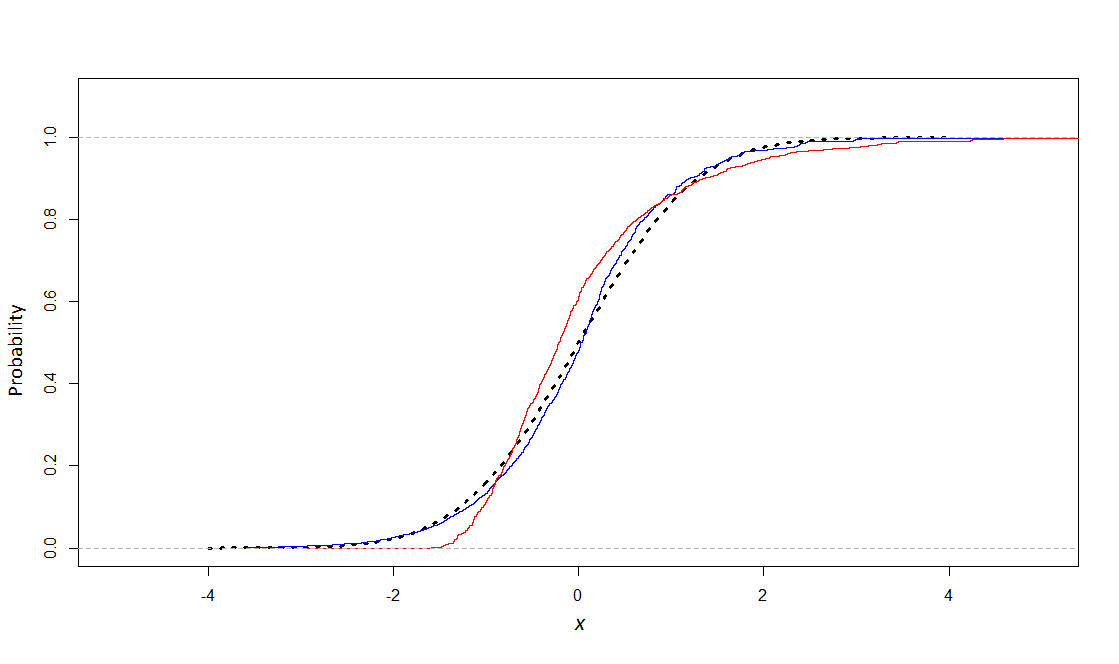}
    \caption{$p=10$}
  \end{subfigure}
  \begin{subfigure}[b]{0.45\linewidth}
    \includegraphics[width=\linewidth,height=2.5in]{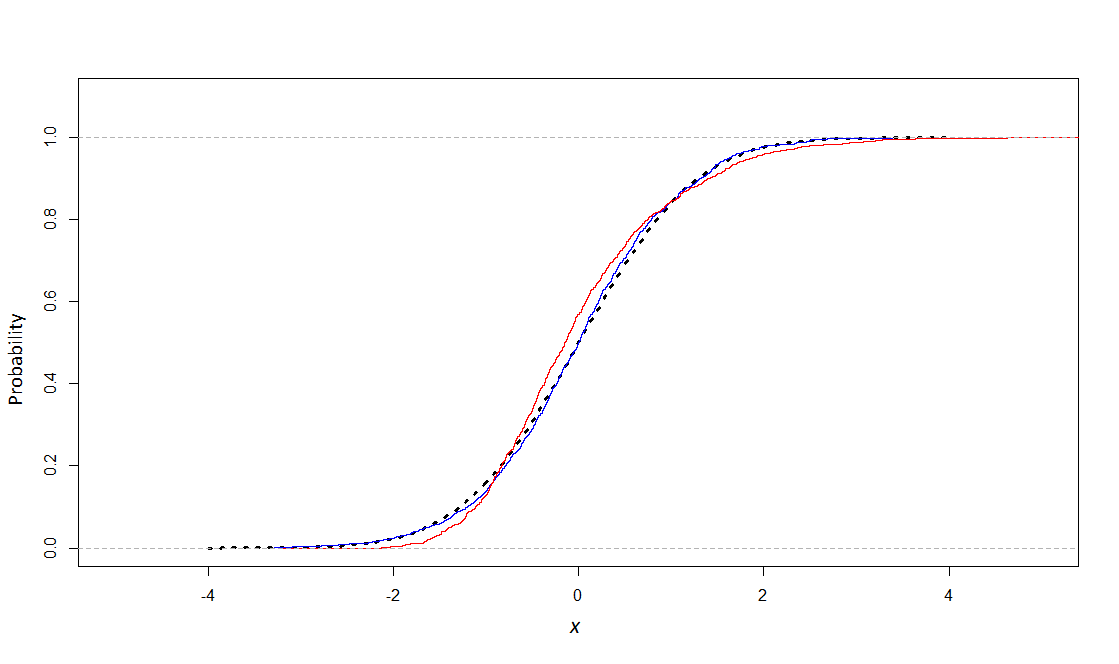}
    \caption{$p=30$}
  \end{subfigure}
    \begin{subfigure}[b]{0.45\linewidth}
    \includegraphics[width=\linewidth,height=2.5in]{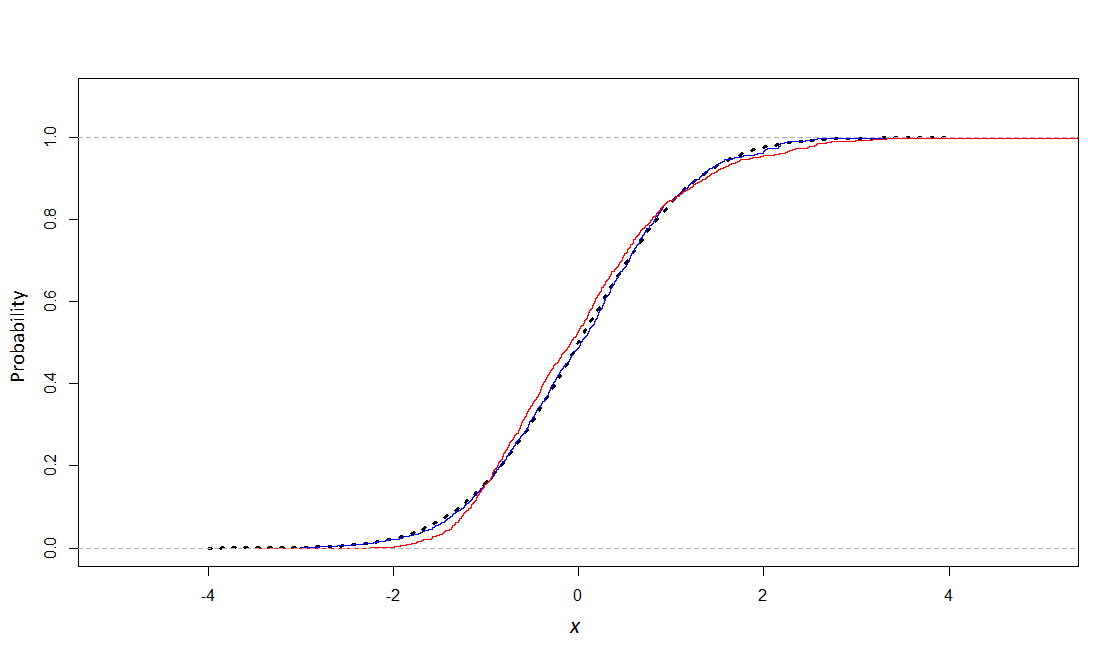}
    \caption{$p=50$}
  \end{subfigure}
      \begin{subfigure}[b]{0.45\linewidth}
    \includegraphics[width=\linewidth,height=2.5in]{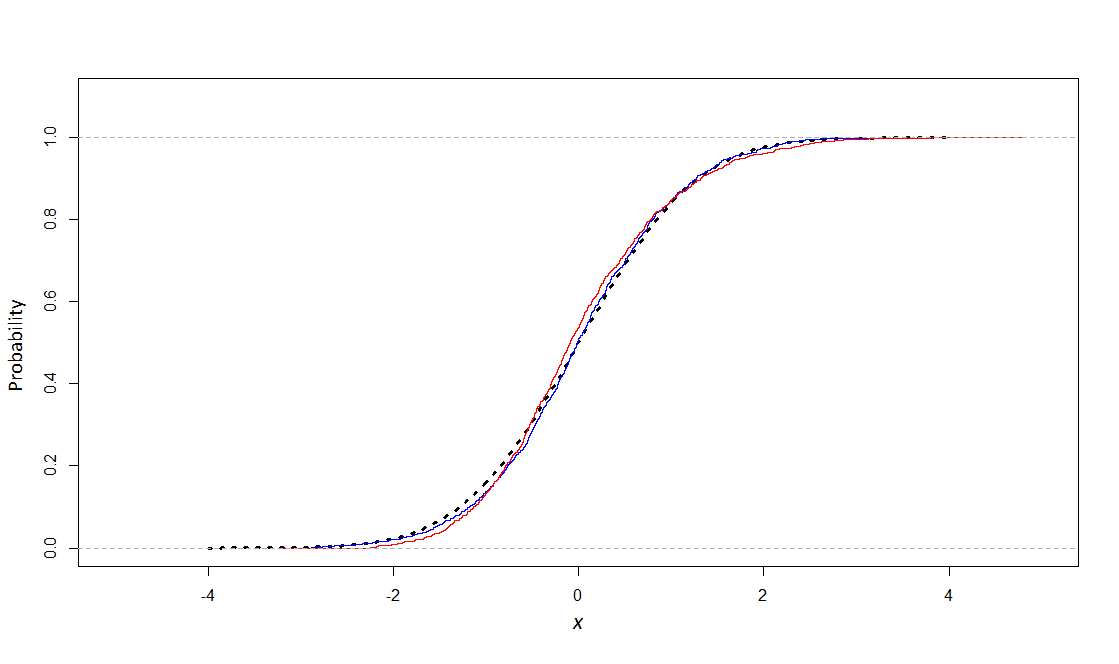}
    \caption{$p=100$}
    \end{subfigure}
        \begin{subfigure}[b]{0.45\linewidth}
    \includegraphics[width=\linewidth,height=2.5in]{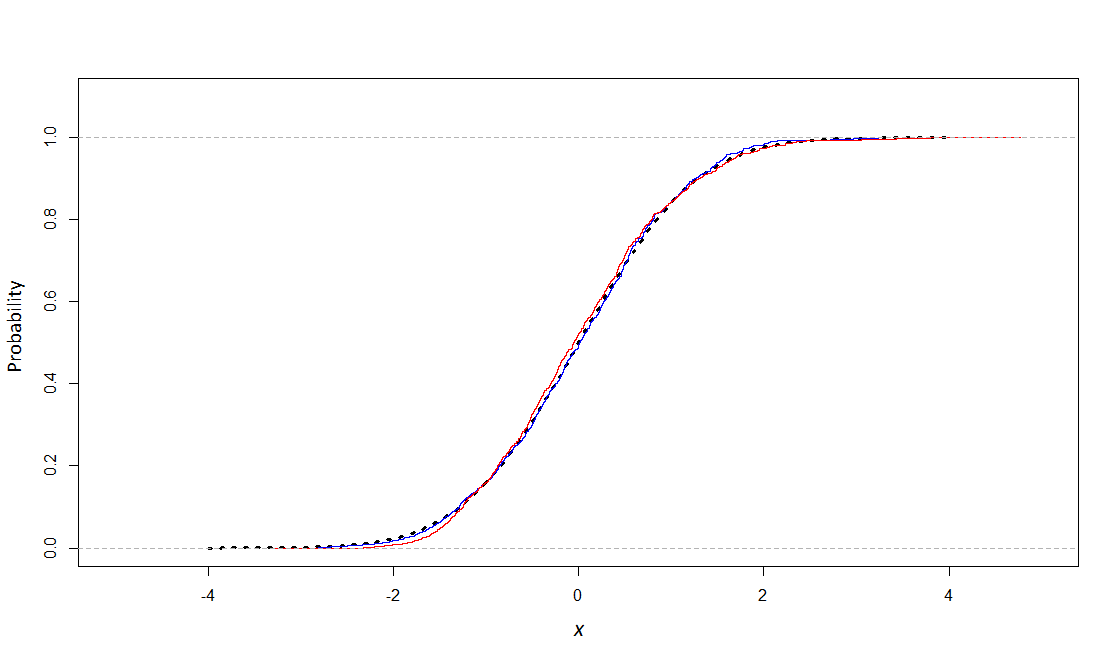}
    \caption{$p=150$}
    \end{subfigure}
        \begin{subfigure}[b]{0.45\linewidth}
    \includegraphics[width=\linewidth,height=2.5in]{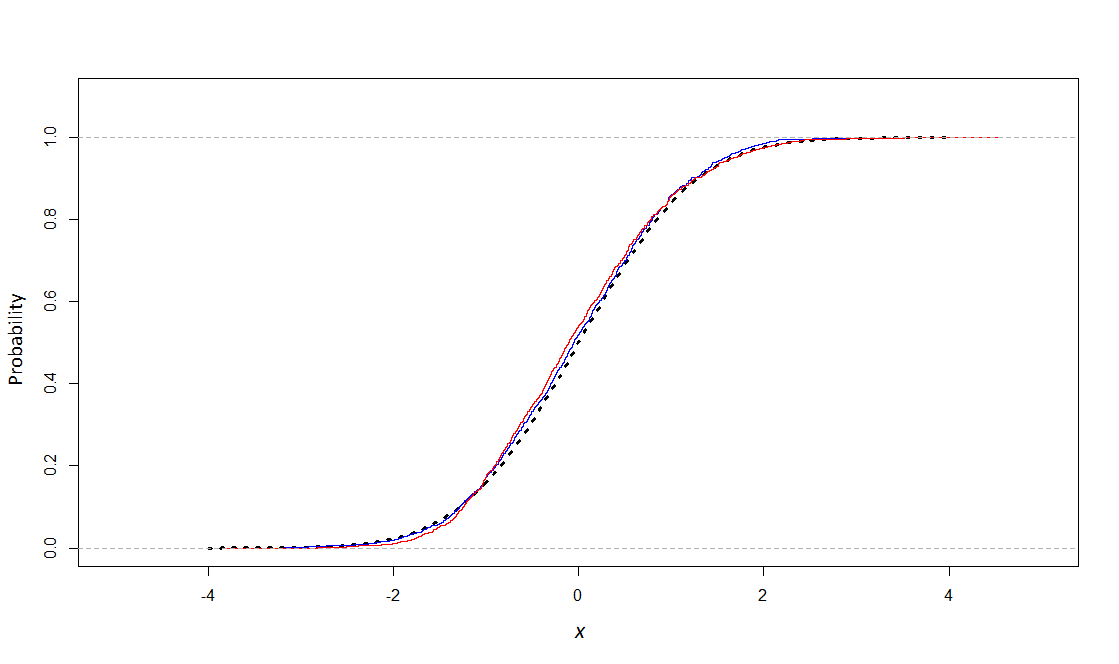}
    \caption{$p=200$}
    \end{subfigure}
  \caption{A comparison of the CDF of the statistic $U_i$ (red-solid curve) and $Z_i$ (blue-solid curve) with standard Normal's CDF (black-dash curve).}
  \label{fig:cdf}
\end{figure}

We first graphically compare the accuracy of the proposed charting statistics $Z_i$ and $U_i$ for varying dimensions $p$. Figure \ref{fig:cdf} compares the cumulative distribution function (CDF) for the statistics $U_i$ and $Z_i$, when the data distributions are the standard multivariate normal, for $p=10, 30, 50, 100, 150, 200$. As we can see, the depicted CDF for the statistic $U_i$ (red curve) generally shows a considerable deviation from the standard normal's CDF (the black-dashed curve), especially in tails and for smaller values of $p$, while the CDF for statistic $Z_i$ (blue curve) is very close to the standard normal. We expected this result as the Cornish-Fisher adjustment provides a better approximation for the quantiles based on the asymptotic normal distribution.

\begin{figure}[h!]
  \centering
  \begin{subfigure}[b]{0.3\linewidth}
    \includegraphics[width=\linewidth,height=1.7in]{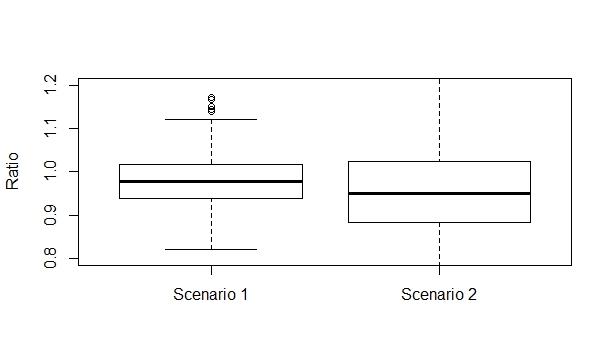}
    \caption{p=30}
  \end{subfigure}
  \begin{subfigure}[b]{0.35\linewidth}
    \includegraphics[width=\linewidth,height=1.7in]{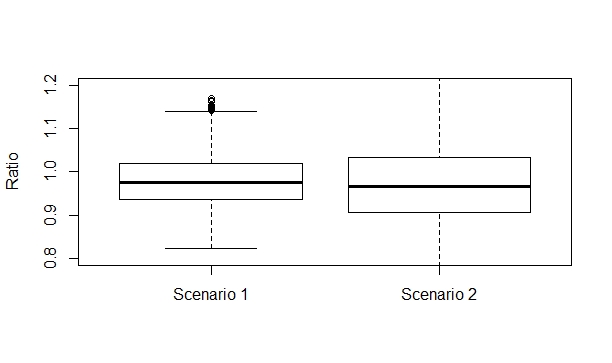}
    \caption{p=50}
  \end{subfigure}
    \begin{subfigure}[b]{0.35\linewidth}
    \includegraphics[width=\linewidth,height=1.7in]{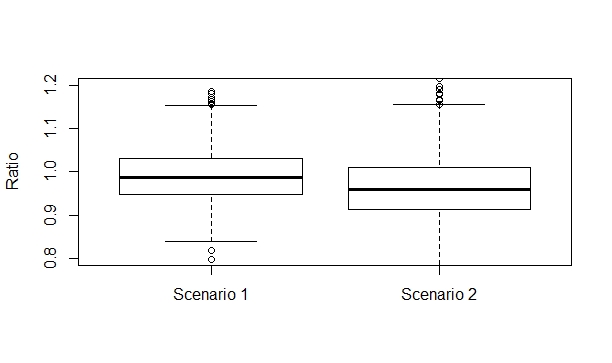}
    \caption{p=100}
  \end{subfigure}
      \begin{subfigure}[b]{0.35\linewidth}
    \includegraphics[width=\linewidth,height=1.7in]{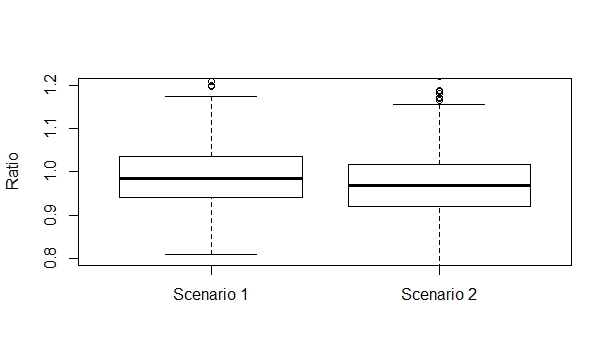}
    \caption{p=150}
    \end{subfigure}
    \begin{subfigure}[b]{0.35\linewidth}
    \includegraphics[width=\linewidth,height=1.7in]{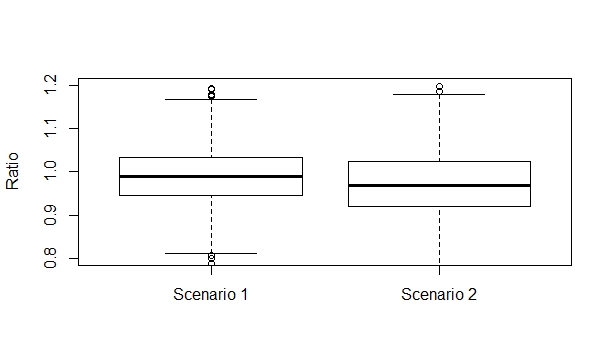}
    \caption{p=200}
    \end{subfigure}
    \begin{subfigure}[b]{0.35\linewidth}
    \includegraphics[width=\linewidth,height=1.7in]{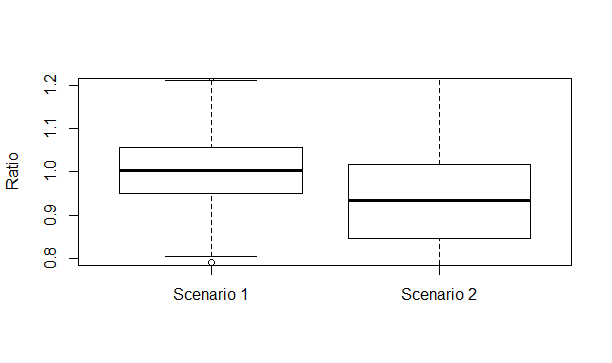}
    \caption{p=30, r=0.1}
    \end{subfigure}
       \begin{subfigure}[b]{0.35\linewidth}
    \includegraphics[width=\linewidth,height=1.7in]{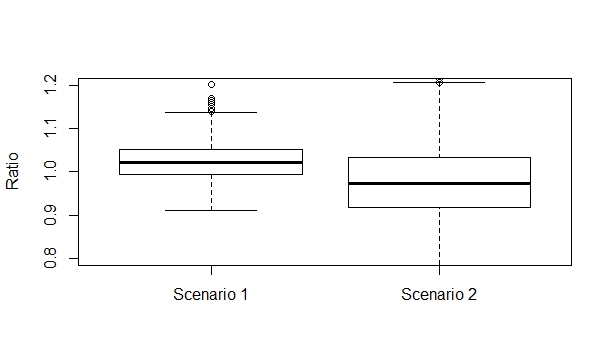}
    \caption{p=30, r=0.2}
    \end{subfigure}
           \begin{subfigure}[b]{0.35\linewidth}
    \includegraphics[width=\linewidth,height=1.7in]{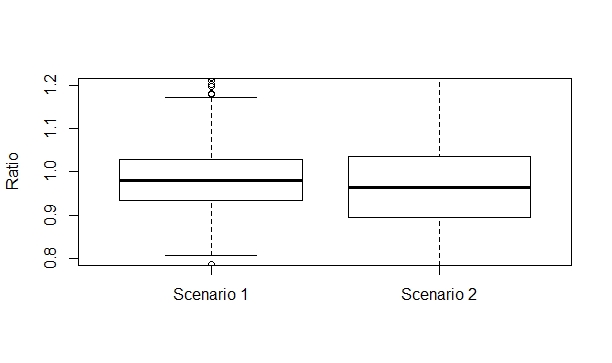}
    \caption{p=100, r=10\%}
    \end{subfigure}
  \caption{Boxplots for the ratio of the re-weighted MDP estimates $\widehat{{\rm tr}(\bfmath{\rho}^3)}_{_{\rm RMDP}}$ to the actual value of ${\rm tr}(\bfmath{\rho}^{3})$ for different $p$ with $m=100$.}
  \label{fig:box}
\end{figure}

To evaluate Phase I performance, we begin by evaluating the estimation accuracy of the proposed robust approach in Section 2.1. Figure \ref{fig:box} depicts the boxplots for the ratio of the re-weighted MDP estimates $\widehat{{\rm tr}(\bfmath{\rho}^3)}_{_{\rm RMDP}}$ to the actual value of ${\rm tr}(\bfmath{\rho}^{3})$ (vertical axis) based on simulations under Scenarios 1 and 2 (horizontal axis). This figure shows the accuracy of the proposed robust estimation method for dimensions $p=30, 50, 80, 100, 150, 200$ and Phase I sample size $m=100$ when there is no contamination in the Phase I data and $\alpha=5\%$ is considered. It also shows the figures when in $r=0.1$ and $0.2$ of observations, half of variables are shifted by amount $\delta=1$. Similar figures can be obtained for other values when contamination is present. From these results, we conclude that for different combinations of $p$, $m$, and $r$ under both Scenarios 1 and 2, the re-weighted MDP estimator $\widehat{{\rm tr}(\bfmath{\rho}^3)}_{_{\rm RMDP}}$ is unbiased for ${\rm tr}(\bfmath{\rho}^3)$. However, the variance under Scenario 1 is generally less than Scenario 2. We expect that we can make similar conclusions for different combinations of $p$ and $m$ under a variety of other scenarios. 

\begin{figure}[h!]
  \centering
  \begin{subfigure}[b]{0.47\linewidth}
    \includegraphics[width=\linewidth, height=2.5in]{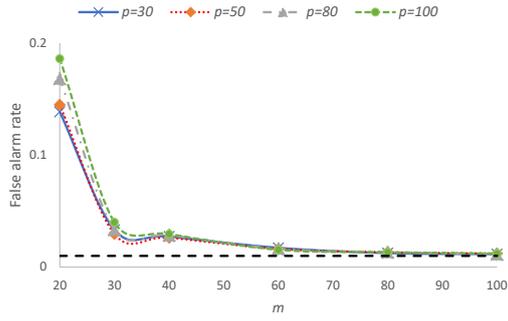}
    \caption{Scenario 1 and $\alpha=0.01$}
  \end{subfigure}
  \begin{subfigure}[b]{0.47\linewidth}
    \includegraphics[width=\linewidth, height=2.5in]{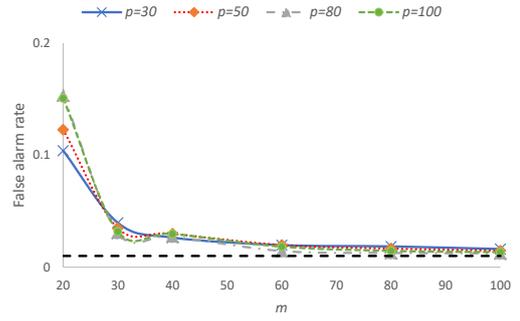}
    \caption{Scenario 2 and $\alpha=0.01$}
  \end{subfigure}
    \begin{subfigure}[b]{0.47\linewidth}
    \includegraphics[width=\linewidth, height=2.5in]{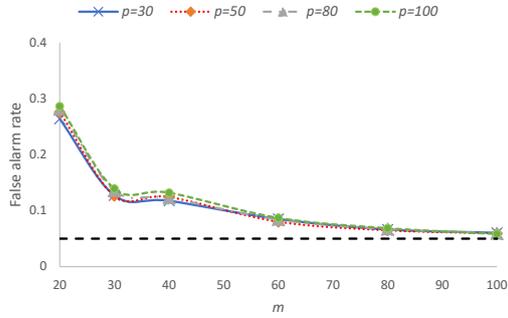}
    \caption{Scenario 1 and $\alpha=0.05$}
  \end{subfigure}
      \begin{subfigure}[b]{0.47\linewidth}
    \includegraphics[width=\linewidth, height=2.5in]{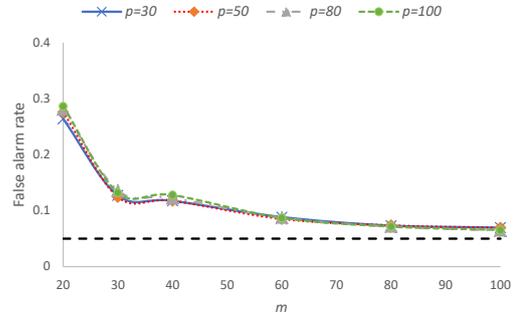}
    \caption{Scenario 2 and $\alpha=0.05$}
    \end{subfigure}
  \caption{Probability of a false alarm for Scenarios 1 and 2 with $p = 30, 50, 80, 100$, where the desired false-alarm rates are 0.01 and 0.05}
  \label{fig:pfa}
\end{figure}

In order to show the statistical performance of our proposed Phase I control chart, we use two criteria. Type I error rate ($\alpha$), i.e., the proportion of
good observations that are wrongly categorized as outliers, and the Type II error rate ($\beta$), which is the
proportion of contaminated observations that are incorrectly labelled as good ones. These error
rates are also called the swamping probability and the masking probability, respectively. Figure \ref{fig:pfa} provides a performance evaluation of the proposed Phase I control chart in achieving the desired false alarm rate when the process is in-control. We consider different values for $m$ and $p$ in our simulations. For each combination of $m$ and $p$, we simulated $k=10,000$ in-control Phase I datasets and recorded the number of out-of-control observations.
The simulated (average) false-alarm rate is then estimated as the average number of out of control points over $k$ simulation runs. Figure \ref{fig:pfa} depicts the estimated  false-alarm rate for $p=30, 100$ and $\alpha=0.01, 0.05$. A reference line for the value of $\alpha$ is also shown as a horizontal dash line. By comparing this figure to Figures 4 and 5 of Williams et al. (2006), it can be concluded that the simulated false-alarm rate of the proposed chart can converge very fast to its nominal value. A comparison is also made in Table \ref{table:alphaCF} between the proposed method and the old one, where CF expansion is used or not in Steps 2 and 3 of Algorithm 1. Table 1 shows that using the CF expansion can considerably enhance the performance of proposed chart in getting the desired Type I error probability.

\begin{table}[h!]

\scriptsize
\centering
\caption{A comparison of simulated $\alpha$ for different values of $p$ and $m=200$, when Algorithm 1 is used with and without CF expansion.}
\label{table:alphaCF}
\begin{tabular}{ |p{0.80cm}|p{0.8cm}|p{0.8cm}|p{0.6cm}|p{0.8cm}|p{0.6cm}|p{0.8cm}|p{0.6cm}|p{0.8cm}|}
 \hline
&\multicolumn{4} {|c|} {\bf Scenario 1}&\multicolumn{4} {|c|}{\bf Scenario 2}\\ \hline
&\multicolumn{2}{|c|}{$\alpha$=0.01} & \multicolumn{2}{|c|}{$\alpha$=0.05}&\multicolumn{2}{|c|}{$\alpha$=0.01} & \multicolumn{2}{|c|}{$\alpha$=0.05}\\ 
\hline 
&  {\centering{\tiny With}} & {\centering{\tiny Without}}& {\tiny With} & {\tiny Without}&  {\centering{\tiny With}} & {\centering{\tiny Without}}& {\tiny With} & {\tiny Without}\\
 &  {\centering{\tiny CF}} & {\centering{\tiny CF}}& {\tiny CF} & {\tiny CF} &  {\centering{\tiny CF}} & {\centering{\tiny CF}}& {\tiny CF} & {\tiny CF}\\\hline
$p$=30& 0.010&\centering{0.022} &0.052&\centering{0.071}&0.013&\centering{0.036} &0.060&0.091\\
$p$=50&0.0095&\centering{0.018} & 0.050&\centering{0.065}& 0.012&\centering{0.029} & 0.058&0.082\\  
$p$=80&0.009&\centering{0.016}  &0.048 &\centering{0.059}&0.011&\centering{0.024}  &0.055 &0.075\\  
$p$=100&0.009 &\centering{0.015} &0.047 &\centering{0.057}&0.011 &\centering{0.022} &0.053 &0.071\\ \hline 
\end{tabular}
\end{table}
Recall, Assumption \ref{A1} implies that correlations among variables are not cumulatively large. This requirement may seem to limit the range of applications for the proposed methodology. However, note that although in the numerator of the proposed statistic only the diagonal elements of the covariance matrix have been involved, the correlation structure is considered in the denominator through ${\rm tr}(\rho^2)$. Besides, our Cornish-Fisher expansion uses ${\rm tr}(\rho^2)$ and ${\rm tr}(\rho^3)$, which directly depend on correlation structure. Indeed, this is one of the reasons that our method can perform better than the similar statistics without CF expansion (look at Table \ref{table:alphaCF} and  Figure \ref{fig:pfa}.) Moreover, very high correlations may happen in applications when dimensionality $p$ is small, but for high dimensional cases it seems unlikely. Hence, it is also useful to consider the effect of correlation on the proposed method. To perform some sensitivity analysis, we provide Figure \ref{fig:sens} and Table \ref{tab1} (in Appendix B) to give better insights into our proposed methods. Figure \ref{fig:sens} investigates the effect of correlation on the simulated false alarm when covariance matrix $\sigma_{ij}=(a)^{|i-j|}\quad {\rm for}\quad i,j=1,\dots,p$ is used and gradually change the value of $a$ from 0 to 0.9. We set the mean vector $\bfmath{\mu}$ to zero. Figure \ref{fig:sens} shows the result for $p=30, 100$ and $\alpha=0.05$, while similar conclusions can be obtained for other values of $p$. It can be concluded from this figure that for small and moderate correlations, the convergence to the nominal false alarm rate is fast, while for some high correlations such as $a=0.9$, the convergence speed is slower. However, we notice that since $M_{_i}(\bfmath{\mu},\,\mathbf{D})$ is a weighted sum of independent $\chi^2_{(1)}$ random variables, we can overcome this shortcoming by adopting the  Welch-Satterthwaite (W-S) $\chi^2$-approximation. Satterthwaite (1941, 1946) and Welch (1947) provided a moment matching approach to approximate the distribution of a weighted sum of independent $\chi^2$ random variables, which statisticians have successfully used over the past 70 years. We postpone a careful investigation of this promising line of research in connection to our work for future research. Nevertheless, we point out that Zhang et al. (2020) recently showed the effectiveness of the Welch-Satterthwaite approximation in their proposed high-dimensional two-sample test statistic when the variables are highly correlated.

\begin{figure}[h!]
  \centering
  \begin{subfigure}[b]{0.45\linewidth}
    \includegraphics[width=\linewidth, height=2in]{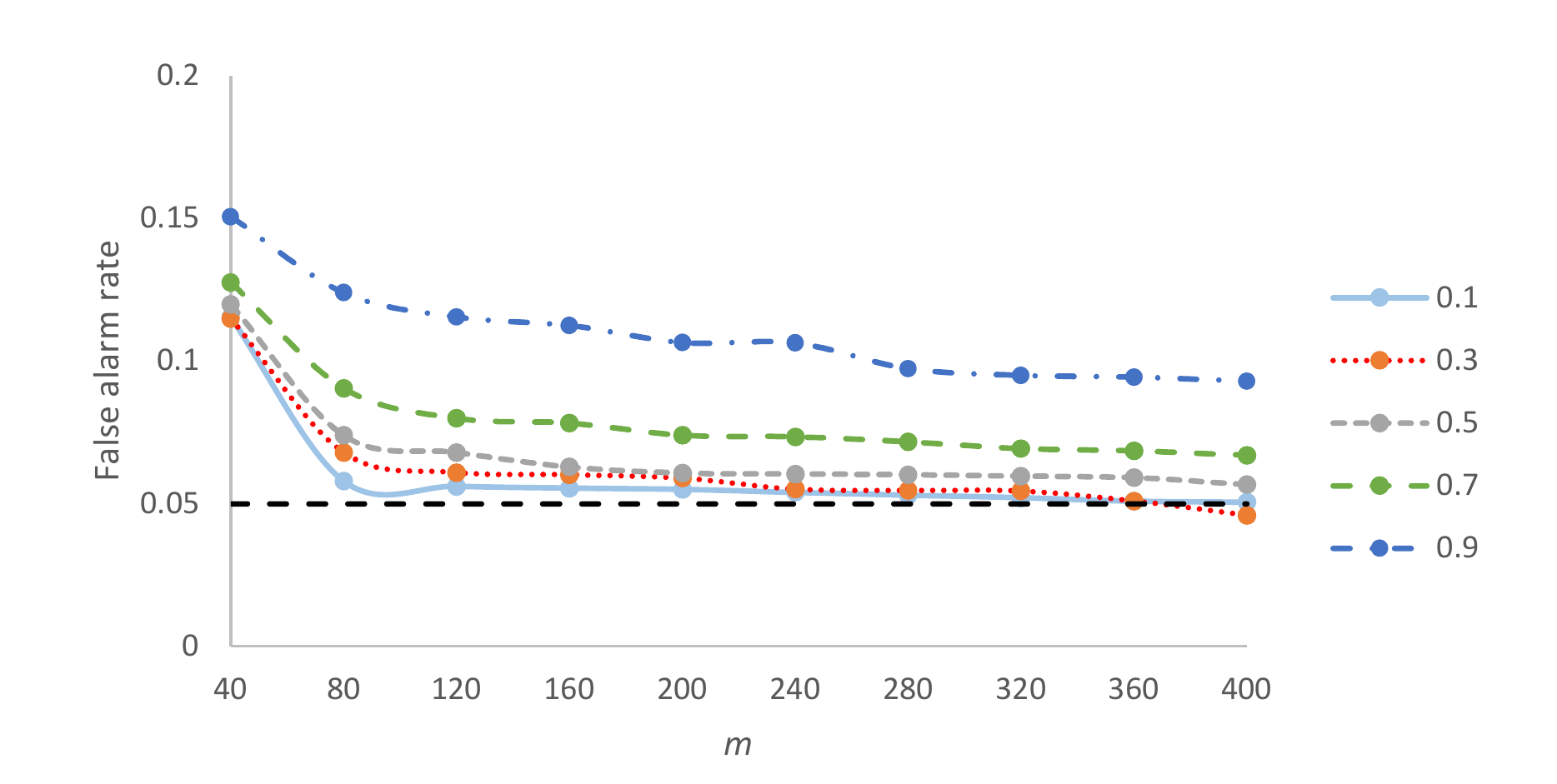}
    \caption{$p=30, \alpha=0.05$ }
  \end{subfigure}
  \begin{subfigure}[b]{0.45\linewidth}
    \includegraphics[width=\linewidth, height=2in]{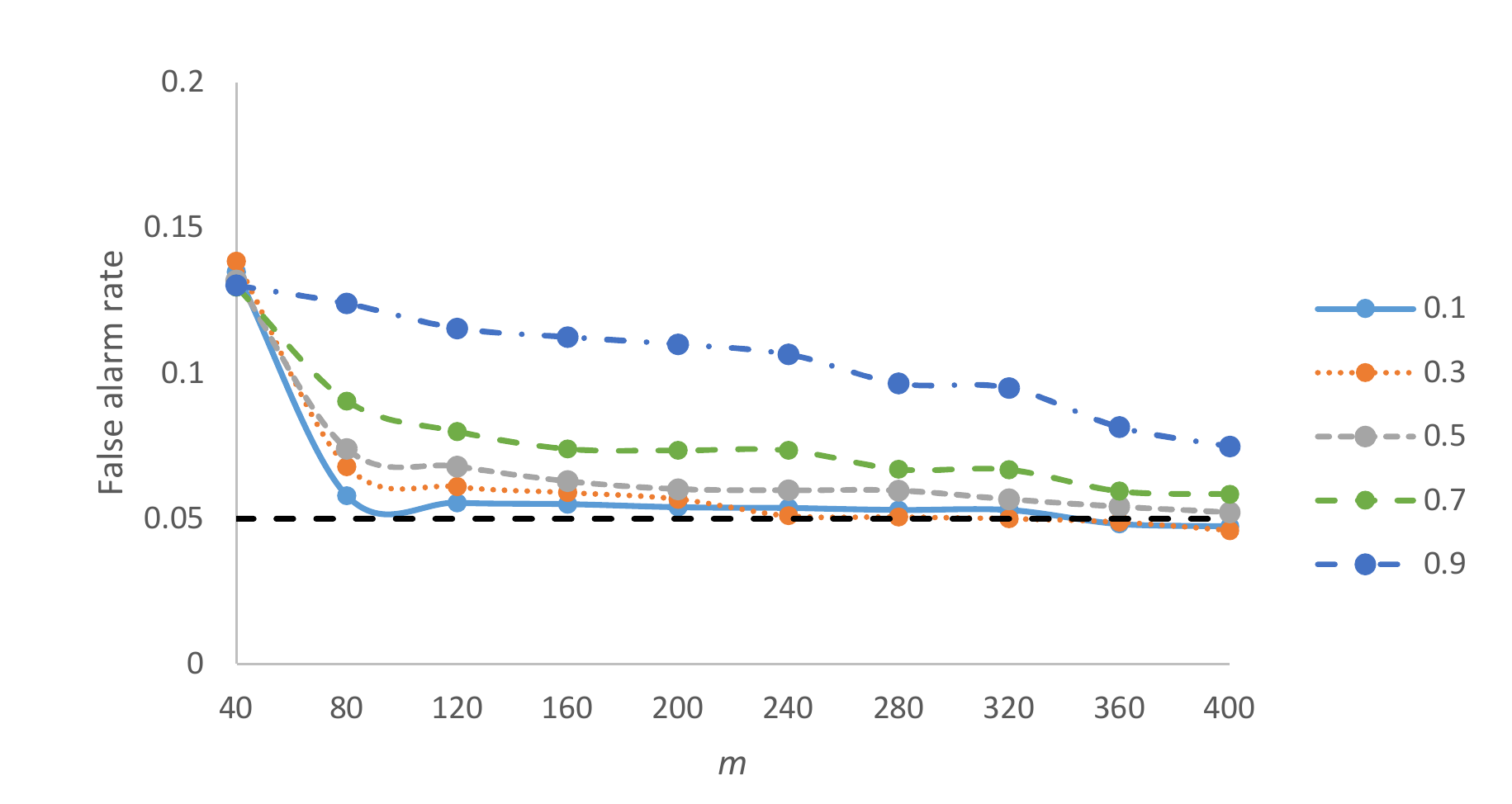}
    \caption{p=100, $\alpha=0.05$ }
  \end{subfigure}
  \caption{Effect of correlation on the false alarm rate of the proposed chart (cavariance matrix $\sigma_{ij}=(a)^{|i-j|}\quad {\rm for}\quad i,j=1,\dots,p$ with different values of $a$) for Scenario 2}
  \label{fig:sens}
\end{figure}

\begin{figure}[h!]
  \centering
  \begin{subfigure}[b]{0.45\linewidth}
    \includegraphics[width=\linewidth, height=2.5in]{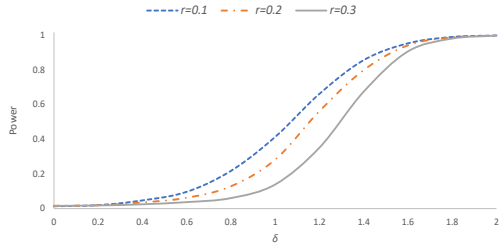}
    \caption{$p=30$ and $\alpha=0.01$}
  \end{subfigure}
  \begin{subfigure}[b]{0.45\linewidth}
    \includegraphics[width=\linewidth, height=2.5in]{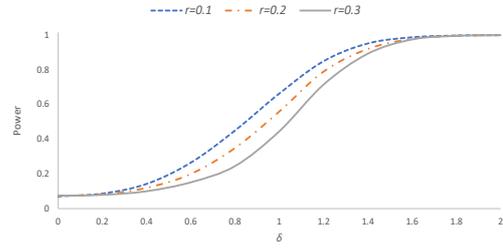}
    \caption{$p=30$ and $\alpha=0.05$}
  \end{subfigure}
    \begin{subfigure}[b]{0.45\linewidth}
    \includegraphics[width=\linewidth, height=2.5in]{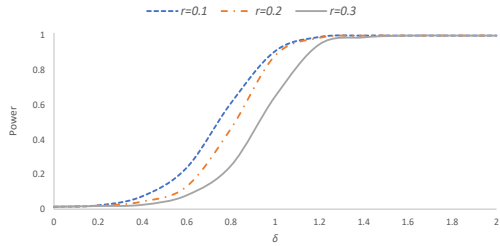}
    \caption{$p=100$ and $\alpha=0.01$}
  \end{subfigure}
      \begin{subfigure}[b]{0.45\linewidth}
    \includegraphics[width=\linewidth, height=2.5in]{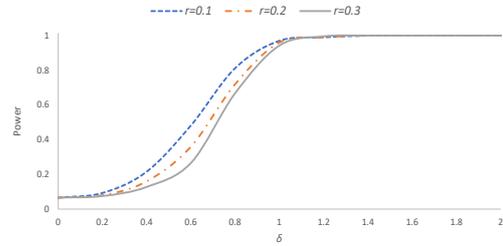}
    \caption{$p=100$ and $\alpha=0.05$}
    \end{subfigure}
  \caption{Power of the proposed chart for Scenario 2 and $p = 30, \,100$, $r=0.1,\, 0.2, \,0.3$, where the desired false-alarm rates are 0.01 and 0.05.}
  \label{fig6}
\end{figure}

We also evaluate the performance of the proposed chart in detecting a mean shift in Phase I for different values of the rate of contamination $r$ and size of shift. We assume that there are $m\times r$ outlying observations with distribution $N(\bfmath{\mu}_{_1},\,\bfmath{\Sigma})$, while $m\times{(1-r)}$ observations are generated from the in-control distribution $N(\bfmath{\mu},\bfmath{\Sigma})$. 
We define $\bfmath{\mu}_{_1}-\bfmath{\mu}=\bfmath{\delta}$, and for simplicity, we assume that all variables of outlying observations are equally shifted by amount  $\delta$ and then report the simulation results in terms of $\delta$. Figure \ref{fig6} depicts the power ($1-\beta$) of propsed chart under Scenario 2 for dimensions $p=30,\, 100$, contamination rates $r=0.1,\, 0.2 ,\, 0.3$, type I errors $\alpha=0.01, 0.05$, and $\delta=0,0.2,0.4, ..., 2$. From figure \ref{fig6}, one can see that when the value of
$\delta$ is zero, the power of control chart is about $\alpha$, while as the value of $\delta$ increases, the power also increases. Moreover, similar to the Phase I control charts based on RMCD (See for example Variyath and Vattathoor 2014, Figures 4-9), we observe that as the contamination rate $r$ increases, the power of proposed RMDP chart decreases.

Finally, we compare our outlier detection method with the popular robust method, the minimum covariance determinant (MCD), which is used by Vargas (2003) for Phase I analysis of multivariate processes with individual observations. Vargas (2003) proposed replacing the classical estimators of mean and covariance with MCD estimators in the $T^2$ statistic and used the following UCL for Phase I control chart

\begin{align}
 \frac{(m-1)^2}{m}B_{\alpha, \frac{p}{2}, \frac{m-p-1}{2}}
 \end{align}

\begin{figure}[h!]
  \centering
  \begin{subfigure}[b]{0.45\linewidth}
    \includegraphics[width=\linewidth, height=1.8in]{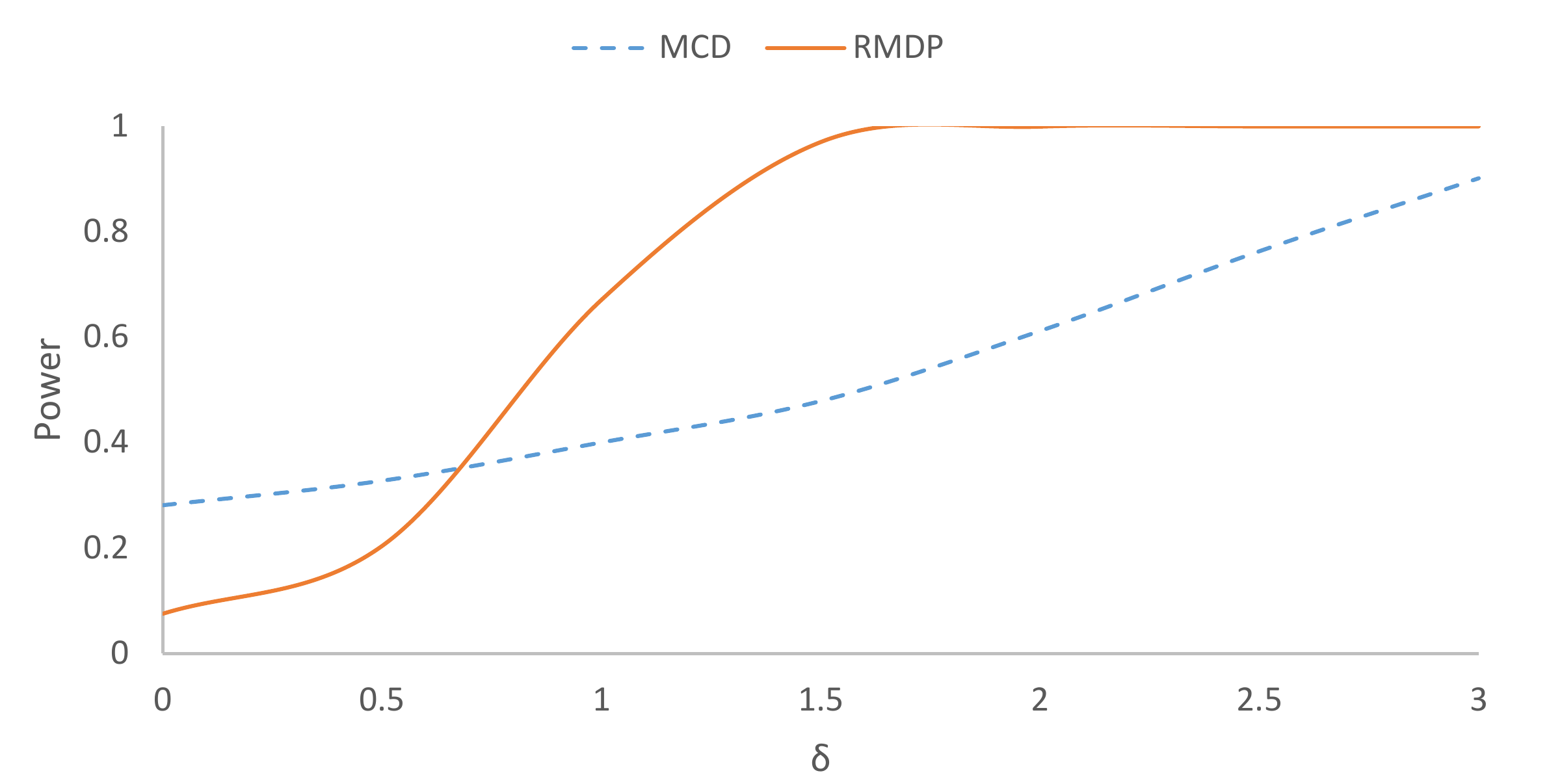}
    \caption{$p=30, m=100$ }
  \end{subfigure}
  \begin{subfigure}[b]{0.45\linewidth}
    \includegraphics[width=\linewidth, height=1.8in]{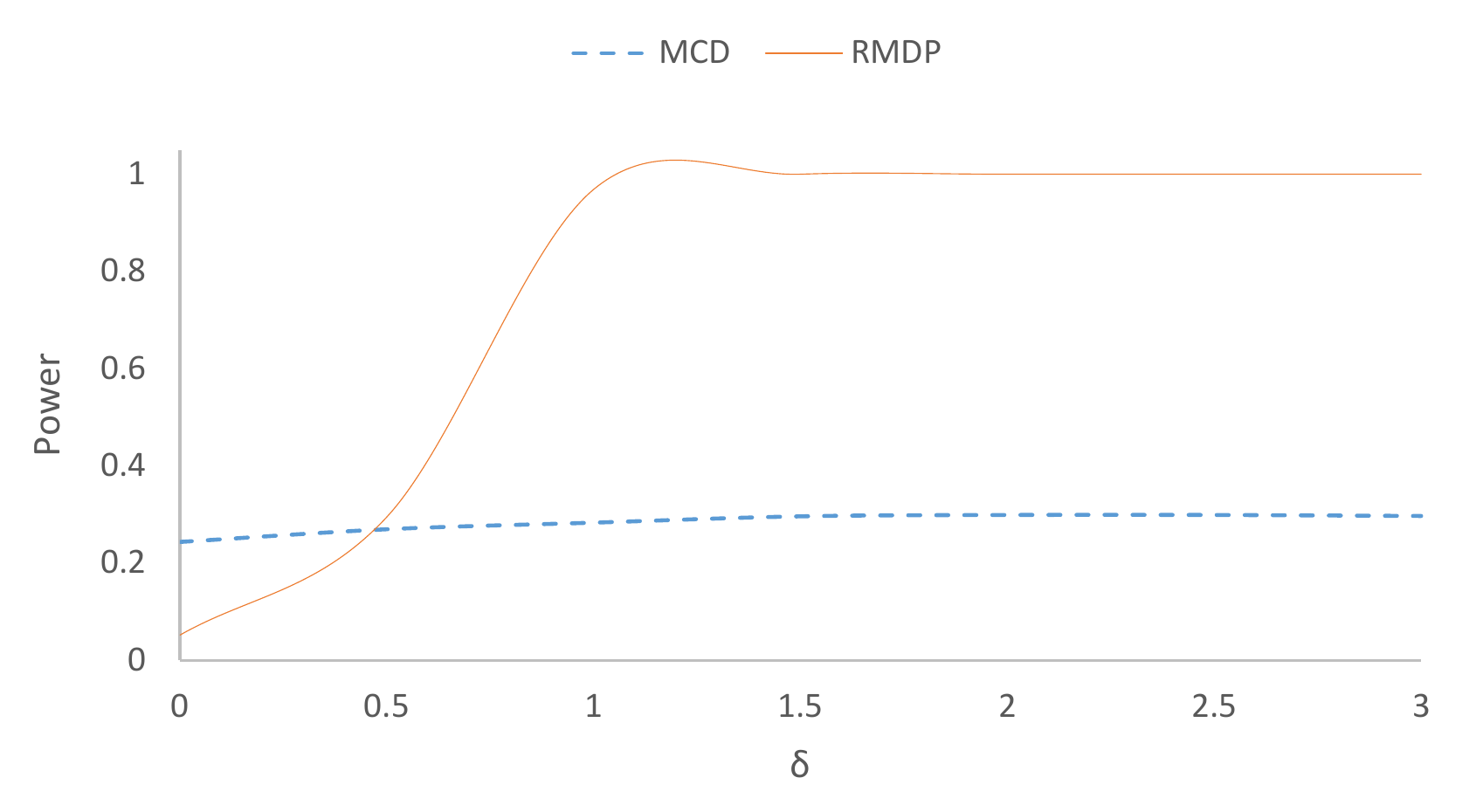}
    \caption{$p=100, m=200$ }
  \end{subfigure}
  \caption{A comparison between the power of the proposed RMDP chart and the MCD chart of Vargas (2003) in Phase I when 10\% of variables are shifted in an amount of $\delta$ in Phase I and the desired $\alpha$ is 0.05.}
  \label{fig7}
\end{figure}

We use the function {\tt CovMcd} in the {\tt rrcov} package of {\tt R} software to calculate process parameters based on MCD in Phase I. Again, out of the generated $m$ observations, $\lfloor m\,r\rfloor$ of them are outliers with distribution ${\rm N}(\bfmath{\mu}_{_1},\,\bfmath{\Sigma})$ and the remaining $\lfloor m\,(1-r)\rfloor$ observations are generated from the in-control distribution ${\rm N}(\bfmath{\mu},\,\bfmath{\Sigma})$. We consider three different rates of contamination $r$ in Phase I data and assume that the mean of the shifted variables are equally shifted by the amount of $\delta=0.5, \,1, \,\,\dots ,\,3$ while the covariance matrix remains in-control. We only report the comparison between the two methods for Scenario 2 when $r=0.1$ and $p=30, 100$, while we can make similar conclusions for Scenario 1. In the Type I comparisons, our RMDP method showed better convergence to the nominal rate. For example, for both $p=30, 100$ and $\alpha=0.05$, the Type I error of our proposed RMDP chart is close to 0.05, while it is almost 0.25 for the MCD chart of Vargas (2003). Figure \ref{fig7} depicts the comparison of (outlier) detection power of the two methods when the nominal significance level $\alpha$ is chosen to be 0.05. From Figure \ref{fig7}, one can conclude that the detection power for both methods increases as $\delta$ increases, but the power of our proposed method is generally better than the MCD method. This superiority is more considerable for the larger value of $p=100$, as our method obtains the highest power much faster than the MCD method.

\section{Real data applications}
This section applies the proposed methodology on two real datasets. We first carry out the proposed Phase I  approach to investigate the VDP dataset introduced in Section 1. The second example involves data from a diagnostic breast cancer process.  
\subsection{VDP data }
The VDP dataset consists of 24 vertical density profiles of wooden boards, for which 314 density measurements were made along a designated vertical line taken 0.002 inches apart for each profile. Several research articles have discussed the VDP dataset, treating it as a multivariate dataset with 24 observations with a dimensionality of 314. Although in the VDP example, we deal with profile data with functional structure, multivariate control charts have been extensively used for monitoring them. See, for example, Zhang and Albin (2009) and Nassar et al. (2021). We intentionally chose this example to show the superiority of our proposed method over some other multivariate methods, as well as how well our proposed approach performs when some assumptions are violated. Because the sample size is smaller than the dimensionality, the sample variance-covariance matrix is singular. To overcome this problem, Wang and Jiang (2009) suggested resampling each profile by taking one point out of every 16 points to reduce the total dimension to 20. So each profile has 20 variables left. However, using this dimension reduction approach may ignore important information, and the estimated covariance matrix may be very different from the real
covariance matrix. This estimation problem would be even more difficult if some outliers contaminate the data. However, our proposed method uses all of the available data.

\begin{figure}[h!]
  \centering
  \begin{subfigure}[b]{0.32\linewidth}
    \includegraphics[width=\linewidth, height=2.3in]{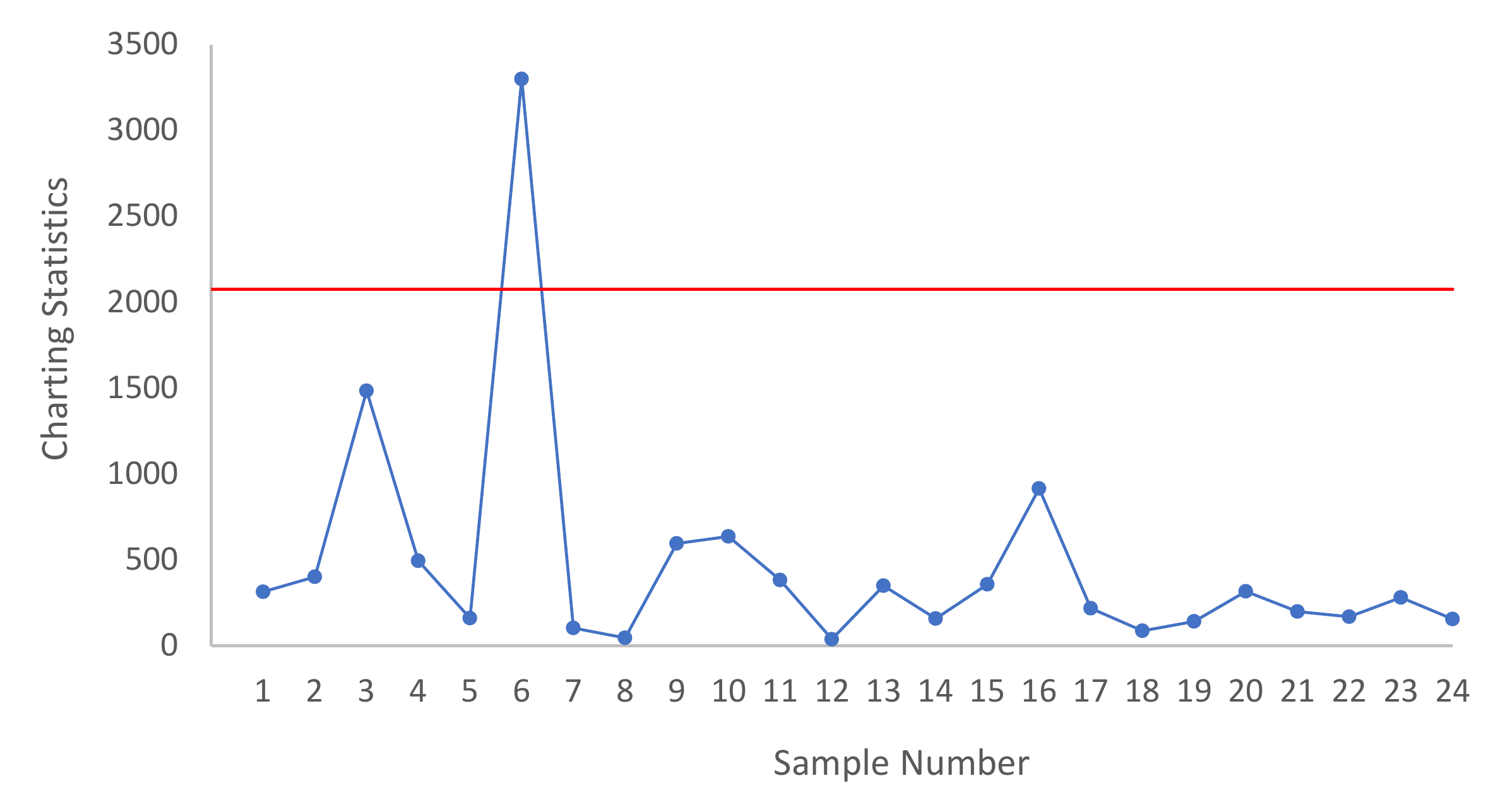}
    \caption{$\alpha$=0.005}
  \end{subfigure}
  \begin{subfigure}[b]{0.32\linewidth}
    \includegraphics[width=\linewidth, height=2.3in]{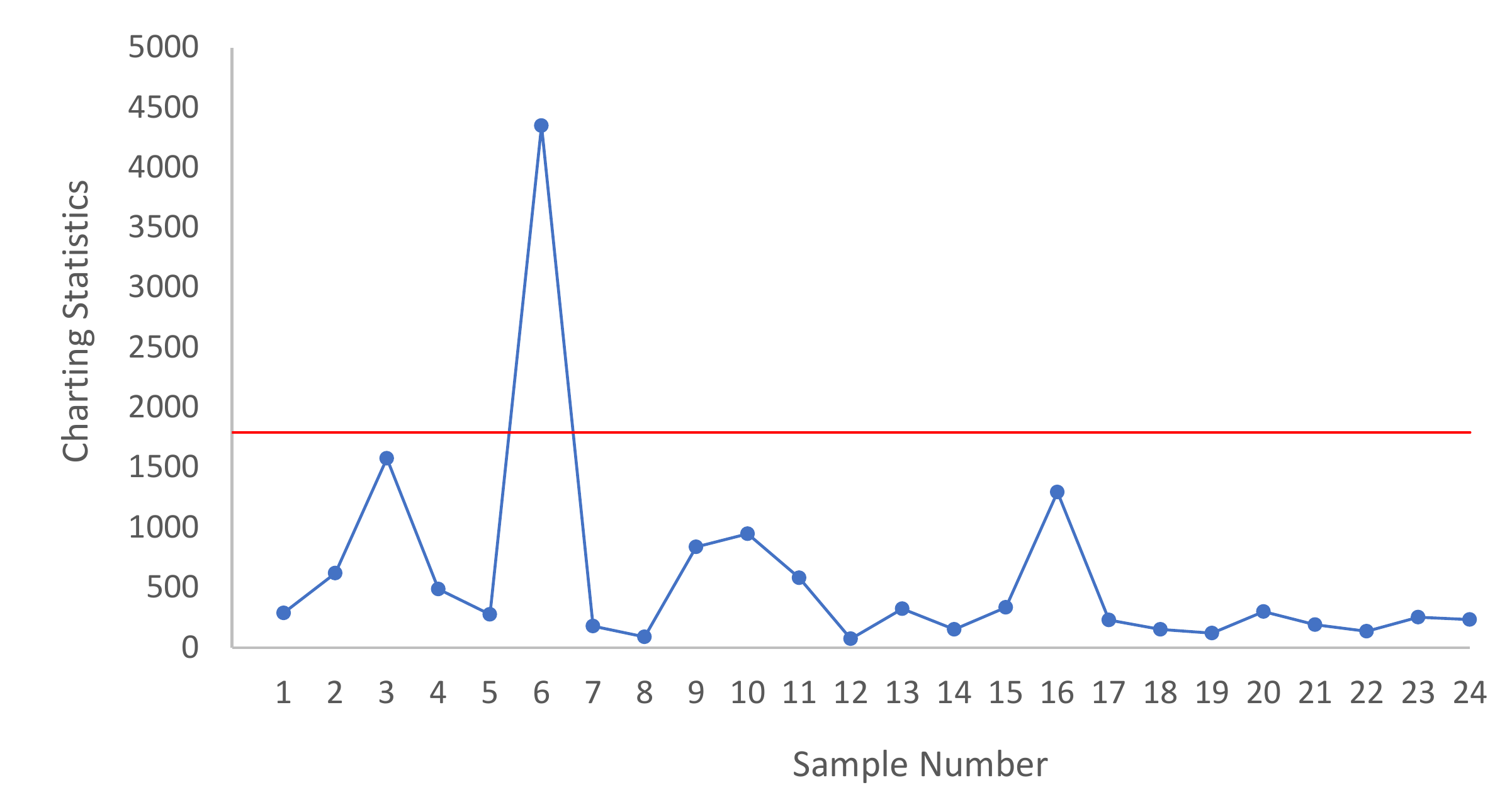}
    \caption{$\alpha$=0.01}
  \end{subfigure}
    \begin{subfigure}[b]{0.32\linewidth}
    \includegraphics[width=\linewidth, height=2.3in]{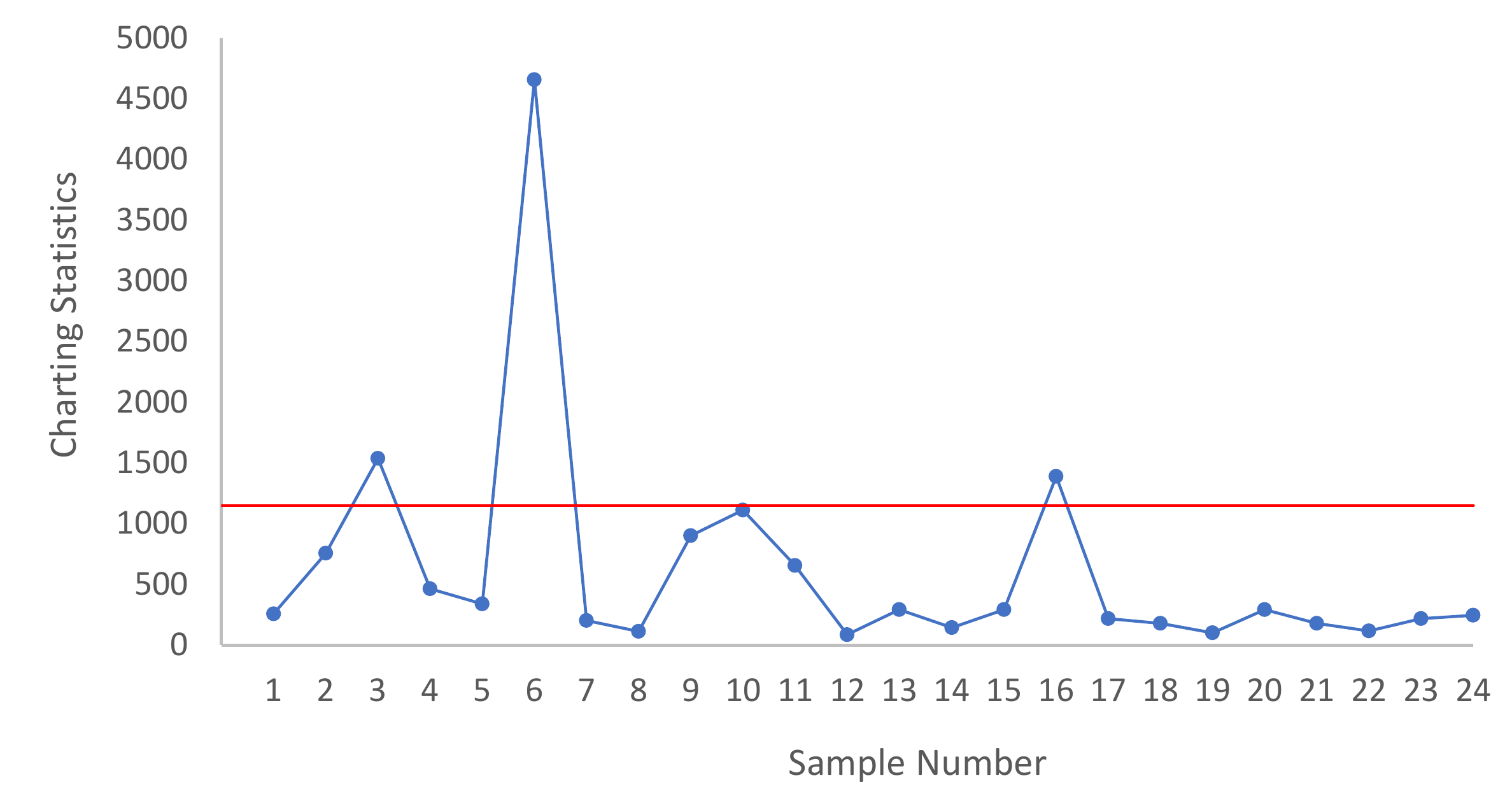}
    \caption{$\alpha$=0.05}
  \end{subfigure}
      
  \caption{Proposed Phase I control chart for VDP data based on different $\alpha$ values.}
  \label{fig8}
\end{figure}

\begin{figure}[h!]
  \centering
    \includegraphics[width=0.9\linewidth]{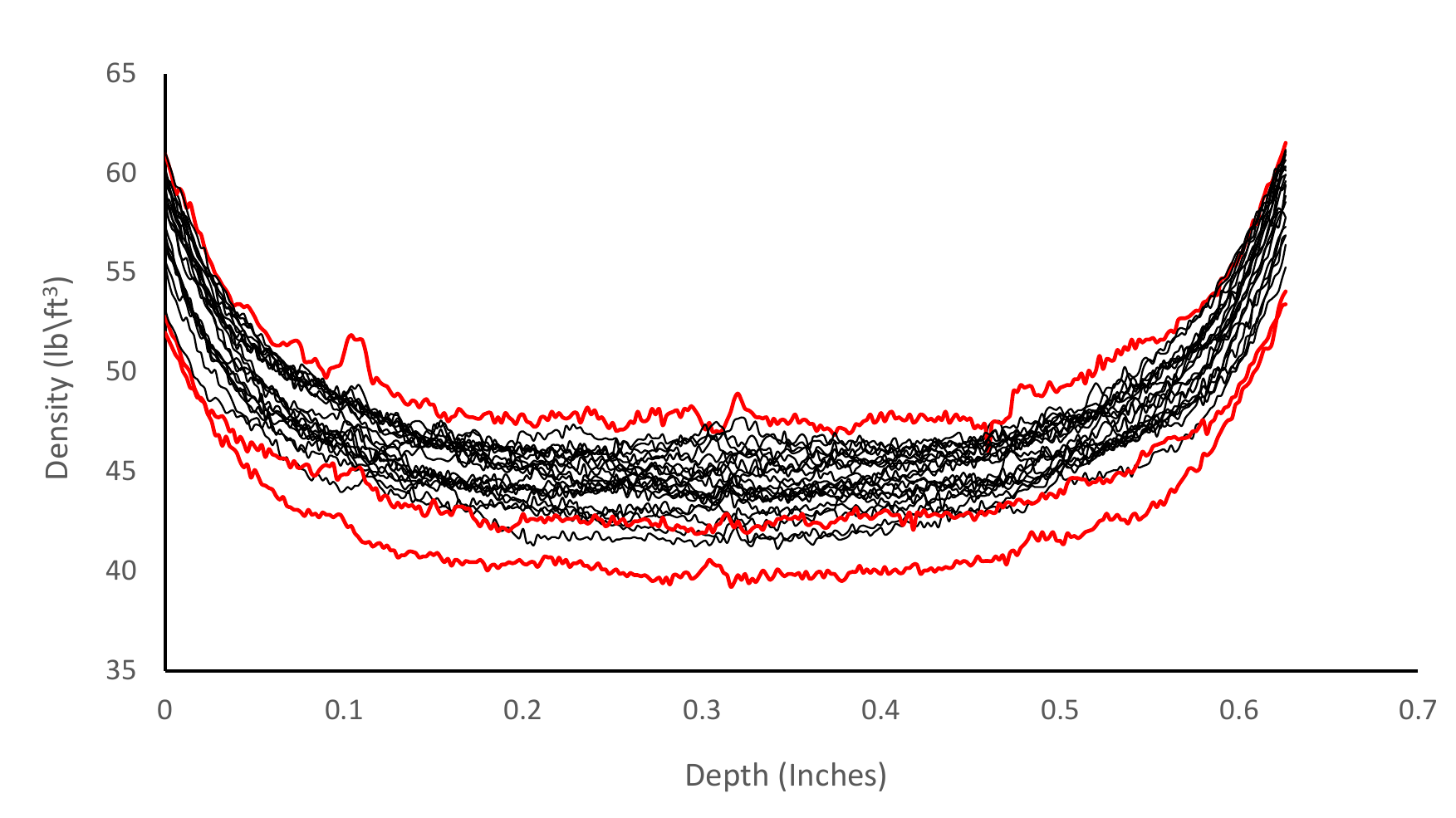}
  \caption{Observed VDP of 24 wooden board profiles with three outlying profiles highlighted in red colour.}
  \label{fig9}
\end{figure}

Wang and Jiang (2009) applied the VS-MSPC chart and Hotelling $T^2$ chart to this data. None of these two charts detected any profile being out of control limits when $\alpha=0.005$. Even for $\alpha=0.01$ and $\alpha=0.05$, $T^2$ chart does not show an out-of-control signal. We consider the VDP data as Phase I data and apply our outlier detection method with $\alpha=0.005, 0.01,$ and 0.05. For these values of $\alpha$, the required parameters are calculated based on 24 profiles and when $\alpha=0.005$, $c^{^{\rm MDP}}_{p,m}=1.11$, $\widehat{{\rm tr}(\bfmath{\rho}^2)}_{_{\rm MDP}}=61,612$, and $\widehat{{\rm tr}(\bfmath{\rho}^3)}_{_{\rm MDP}}=13,756,165$ are estimated. The estimated covariance matrix shows that all variables
are strongly correlated, so as mentioned in the last section, we can still use our proposed method with caution. We plot the values of $M^{2}_{i}(\widetilde{\bfmath{\mu}}, \,\widehat{\mathbf{D}}_{_{\rm RMDP}})$ against the threshold given based on equation \eqref{tresh} to show which observation(s) are assigned a zero weight. The resulting control charts based on three values of $\alpha$ are depicted in Figure \ref{fig8}. Note that based on \eqref{e8}, the value of charting statistic for each sample will change by changing $\alpha$, so we cannot combine these three charts into one chart with three different control limits. As is obvious from this figure, when $\alpha=0.005$, the method detects profile 6 as an outlier. Nevertheless, due to the high correlation, we might have weaker detection power than expected in an out-of-control situation. By comparing the control chart for $\alpha=0.005$ with that for $\alpha=0.01$ and $0.05$, we can conclude that samples 3, 6 and 16 need careful investigation. We highlight these three boards using a red colour in Figure \ref{fig9}, among the 24 observed VDPs profiles. The result of our investigation is consistent with the two outliers (boards 3 and 6) detected by the third method proposed by Williams et al. (2003).  Notice that the method of Williams et al. (2003) is based on Intra-Profile Pooling and assumes that no variability is due to a common cause. 

\subsection{An example of diagnostic breast cancer dataset}

In this subsection, we provide another example using a multivariate dataset of diagnostic breast cancer. The dataset is available through the ftp server in the Computer Sciences Department at UW-Madison: http://ftp.cs.wisc.edu/
math-prog/cpo-dataset/machine-learn/cancer/WDBC/ and has been recently used by Fan et al. (2021) for detecting change in Phase I of the high-dimensional covariance matrix. This dataset consists of  567 vector observations, and for each observation, there are 30 real-valued features/variables measured for each observation, as well as a categorical label to indicate if the underlying observation is  benign or malignant, denoted by "B" and "M", respectively. The dataset doesn't contain any null (missing) values and in total,  357 observations labeled as “B” and 212 observations labeled as “M". Similar to Fan et al. (2021), we treat the 357 benign observations
as Phase-I in-control observations and the 212 malignant observations as Phase-I out-of-control observations and trying to detect whether there is a shift in the mean between these two types. We also place the malignant observations after the benign observations, meaning that observation 357 is the true change point. 
 
We first use Shapiro–Wilks marginal normality tests, ignoring the correlation between variables. As the p-values are very small, we conclude that the assumption of normality does not hold for most of the variables. So, for each marginal observation $X_{ij},\quad i=1,...,30,\quad j=1, ...,567$, we use the inverse transformation $\Phi^{-1}(\widehat{F}_{i}(X_{ij})),$ where $\widehat{F}_{i}$ is the marginal empirical distribution function based on the 357 in-control observations of the $i$th variable. This transformation has been also used by Li et al .(2020) for multistream data.

\begin{figure}[h!]
  \centering
    \begin{subfigure}[b]{1\linewidth}
    \includegraphics[width=\linewidth]{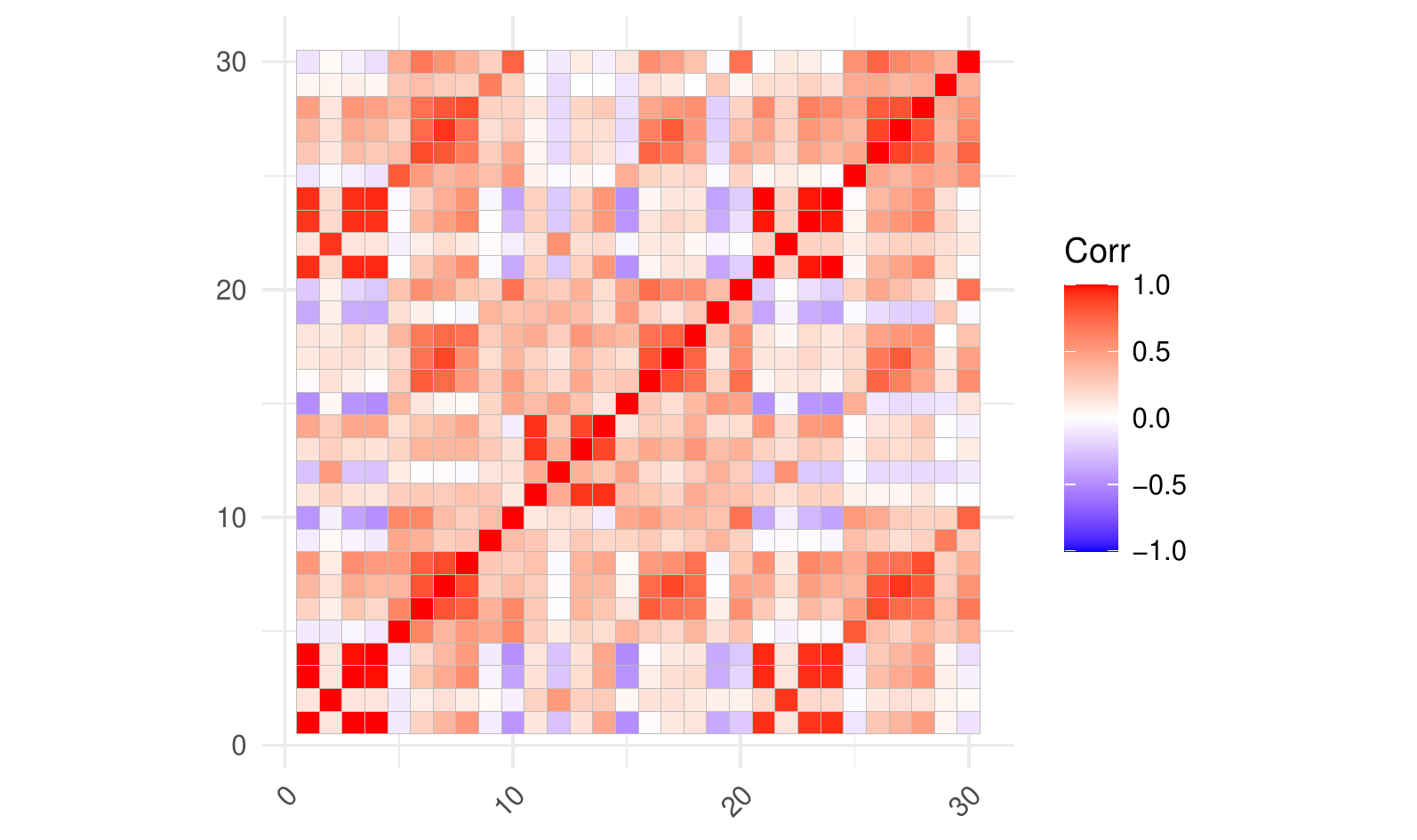}
  \end{subfigure}  
  \caption{Correlation between the 30 variables of diagnostic breast cancer data.}
  \label{fig10}
\end{figure}

\begin{figure}[h!]
  \centering
    \begin{subfigure}[b]{1\linewidth}
    \includegraphics[width=\linewidth]{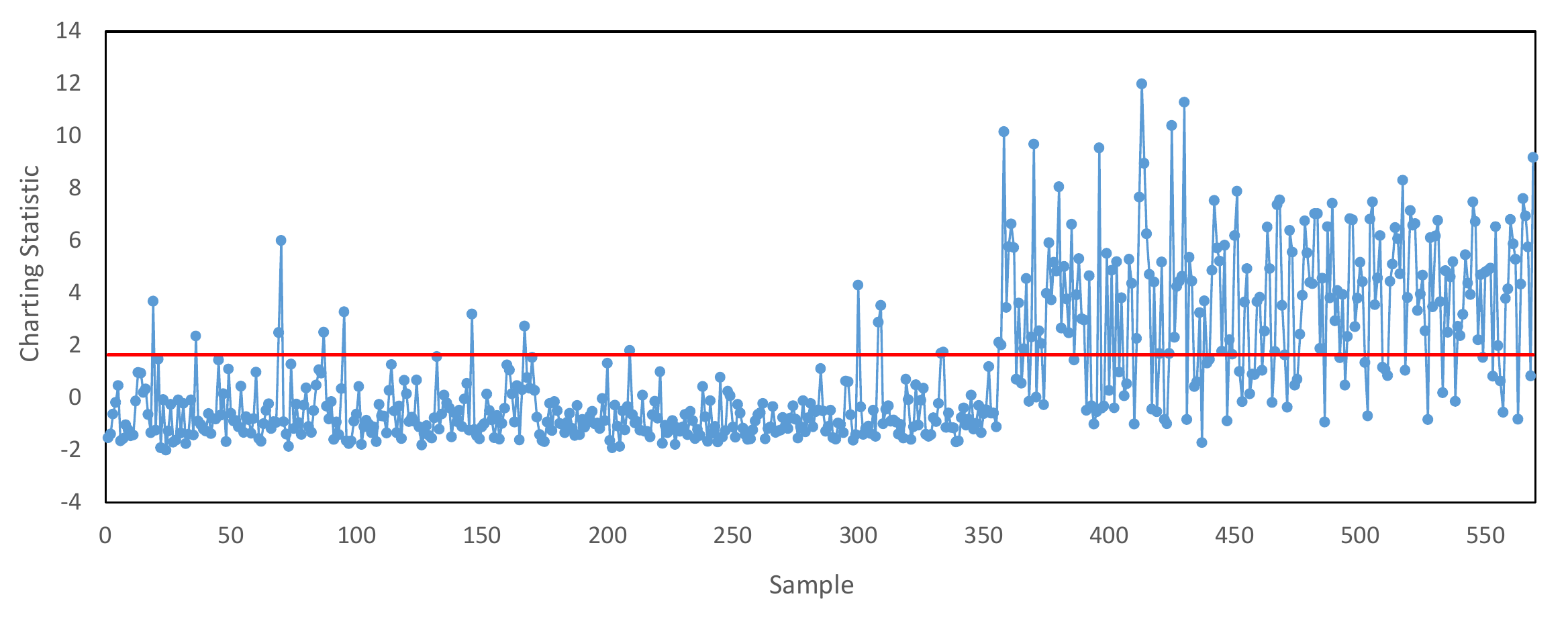}
  \end{subfigure}  
  \caption{A shift in the breast cancer dataset after sample 357 based on proposed RMDP control chart}
  \label{fig11}
\end{figure}

\begin{figure}[h!]
  \centering
    \begin{subfigure}[b]{1\linewidth}
    \includegraphics[width=\linewidth]{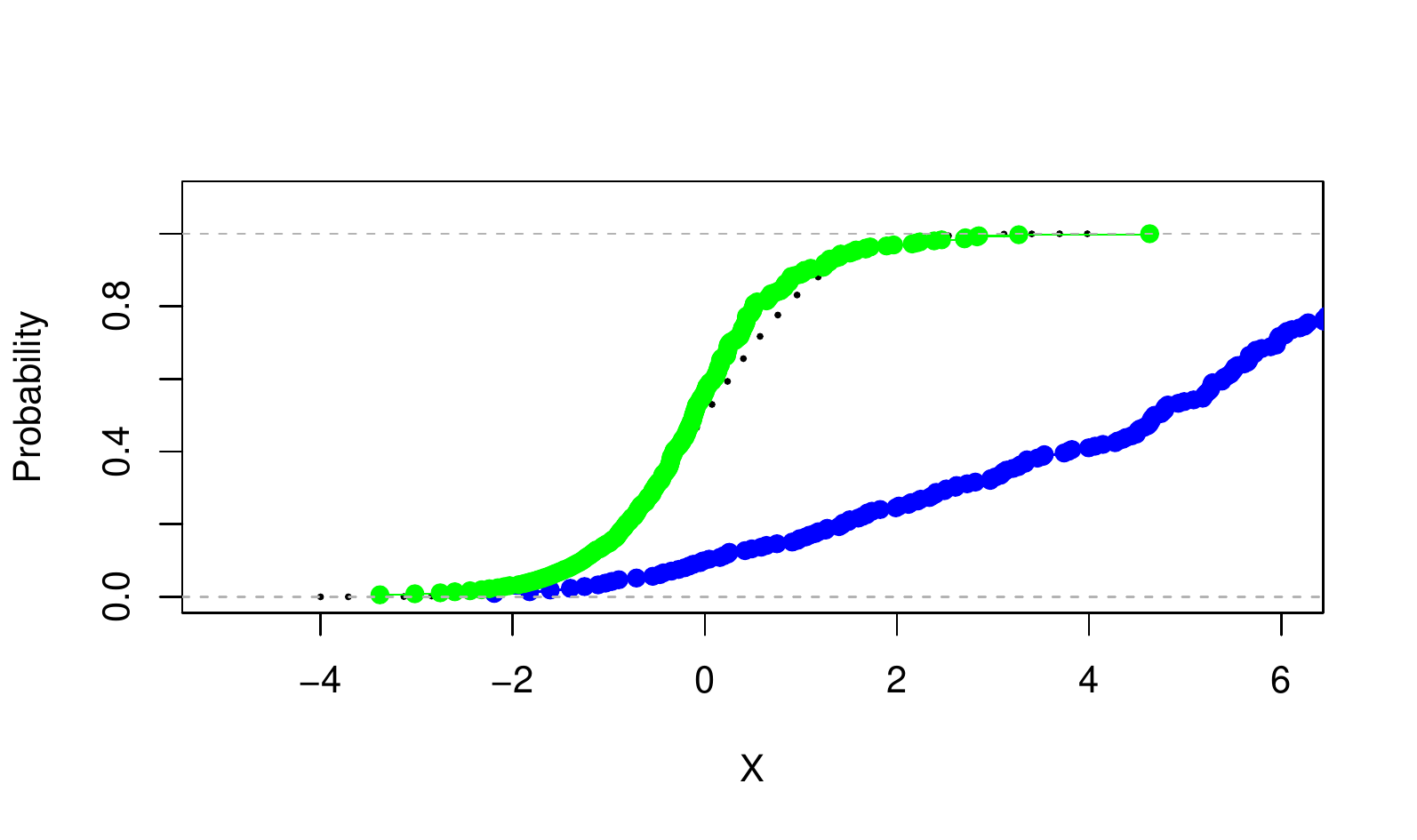}
  \end{subfigure}  
  \caption{A comparison between empirical CDF of charting statistics of bengin observations (green) and malignant observations (blue) with the CDF standard normal (dashed black line).}
  \label{fig12}
\end{figure}

The estimated correlation matrix based on Phase I samples is shown in Figure \ref{fig10}. It can be seen that while a few variables are independent, some of them are (highly) correlated, meaning that high pairwise correlation exists in the dataset. We considered different values of $\alpha$ for this example. The output of our proposed robust RMDP approach on this dataset with $\alpha=0.05$ shows that  $c^{^{\rm MDP}}_{p,m}=1.007$, $\widehat{{\rm tr}(\bfmath{\rho}^2)}_{_{\rm MDP}}=177.6$, and $\widehat{{\rm tr}(\bfmath{\rho}^3)}_{_{\rm MDP}}=1468.7$. Figure \ref{fig11} depicts the control chart based on our proposed method, plotting the statistic in equation \eqref{tresh} against the control limit, where the UCL is $z_{0.05}=1.65$. Figure \ref{fig11} suggests an obvious shift in the process starting from observation 357 in Phase I with a lot of out-of-control samples, which implies that the proposed chart can properly detect and declare the change from bengin to malignant observations as it started right after sample 357.  
As a complementary,  Figure \ref{fig12} compares the empirical CDF of the charting statistics of bengin and malugnant samples with that of the standard normal. While the empirical CDF of conforming samples shows a perfect match with the standard normal CDF, the empirical CDF of malignant items represents a substantial shift to the right from the standard normal distribution, revealing a noticeable change in distribution.

\section{Conclusion}
Technological advancements have led in the development of high-dimensional multivariate processes that require the monitoring of a vast number of variables simultaneously. In such cases, the Phase I sample size is usually small and computing the sample covariance matrix is impractical. This paper employs a new test statistic which is efficient when the sample size in Phase I is less than number of variables. We also propose a consistent estimator for parameter estimation as well as a finite sample correction coefficient via simulation. The robust procedure used in this paper makes the proposed control chart effective in the presence of outlying observations in Phase I. Through Monte Carlo simulation, the proposed procedure shows good performance in Phase I in terms of probability of signal. Altough, Martinez et al. (2020) proposed a bisection algorithm for finding the empirical threshold of RMDP for some non-normal distribution, modifying the control limits for distributions other than normal seems to provide a more general method which is not a function of $m$, $p$, and $\rho$. As future work, we would like to extend the method to more complicated cases such as non-normal high dimensional processes, highly correlated multivariate processes, or when the batch size is greater than one. 

\section*{References}
\begin{enumerate}
\item[] Abdella, G.M., Al‐Khalifa, K.N., Kim, S., Jeong, M.K., Elsayed, E.A. and Hamouda, A.M., 2016. Variable Selection‐based Multivariate Cumulative Sum Control Chart. {\it Quality and Reliability Engineering International, 33}, 565-578.
\item[] Bai, Z., Saranadasa, H. 1996. Effect of high dimension: by an example of a two sample problem. {\it Statistica Sinica, 6,} 311-329.
\item[] Bersimis, S., S. Psarakis, and J. Panaretos. 2007. Multivariate Statistical Process Control Charts: An Overview. {\it Quality and Reliability Engineering International, 23}, 517–543.
\item[] Capizzi, G., Masarotto, G. 2011. A least angle regression control chart for multidimensional data. {\it Technometrics, 53},  285–296.
\item[] Capizzi, G. and Masarotto, G., 2015. Comparison of phase II control charts based on variable selection methods. {\it Frontiers in Statistical Quality Control, 11}, 151-162. Springer International Publishing.
\item[] Chenouri, S. E., Steiner, S. H., \& Variyath, A. M., 2009. A multivariate robust control chart for individual observations. {\it Journal of Quality Technology, 41}, 259-271.
\item[] Croux, C., Haesbroeck, G. 1999. Influence function and efficiency of the minimum covariance determinant scatter matrix estimator. {\it Journal of Multivariate Analysis, 71}, 161-190.
\item[] Ebadi, M., Chenouri, S., Lin, D. K., Steiner, S. H. 2021a. Statistical monitoring of the covariance matrix in multivariate processes: A literature review. {\it Journal of Quality Technology}, 1-136.
\item[] Ebadi, M., Chenouri, S., Steiner, S. H. 2021b. On Monitoring High-Dimensional Multivariate Processes with Individual Observations. {\it Arxiv Preprint}.
\item[] Fan, J., Shu, L., Yang, A.,  Li, Y. (2021). Phase I analysis of high-dimensional covariance matrices based on sparse leading eigenvalues. {\it Journal of Quality Technology, 53}, 333-346.
\item[] Hotelling, H., 1947. {\it Multivariate quality control, illustrated by the air testing of sample bombsights. Techniques of Statistical Analysis.} McGraw Hill, New York, pp. 111–184.
\item[] Jiang, W., Wang, K., Tsung, F. 2012. A variable-selection-based multivariate EWMA chart for process monitoring and diagnosis. {\it Journal of Quality Technology, 44}, 209–230.
\item[] Lehmann, E. L., Romano, J. P. 2006. {\it Testing statistical hypotheses}. Springer Science \& Business Media.
\item[] Li, W., Xiang, D., Tsung, F., Pu, X. 2020. A diagnostic procedure for high-dimensional data streams via missed discovery rate control. {\it Technometrics, 62}, 84-100.

\item[] Martinez, W. G., Weese, M. L., \& Jones-Farmer, L. A. (2020). A one -class peeling method for multivariate outlier detection with applications in phase I SPC.{\it Quality and Reliability Engineering International, 36}, 1272-1295.
\item[]
Nassar, S. H., \& Abdel-Salam, A. S. G. (2021). Semiparametric MEWMA for Phase II profile
monitoring. {\it Quality and Reliability Engineering International, 37}, 1832-1846.
\item[] Pison, G., Van Aelst, S., Willems, G. 2002. Small sample corrections for LTS and MCD. {\it Metrika, 55}, 111-123.
\item[] Ro, K., Zou, C., Wang, Z., \& Yin, G. (2015). Outlier detection for high-dimensional data. {\it Biometrika, 102}, 589-599.
\item[] Rousseeuw, P. J., Van Driessen, K. 1999. A fast algorithm for the minimum covariance determinant estimator. {\it Technometrics, 41}, 212-223.
\item[] Satterthwaite, F. E. 1941. Synthesis of variance. {\it Psychometrika, 6}, 309-316.
\item[] Satterthwaite, F. E. 1946. An approximate distribution of estimates of variance components. {\it Biometrics bulletin, 2}, 110-114.
\item[] Srivastava, M. S., 2005. Some tests concerning the covariance matrix in high dimensional data. {\it Journal of the Japan Statistical Society, 35}, 251-272.
\item[] Srivastava, M. S. 2009. A test for the mean vector with fewer observations than the dimension under non-normality. {\it Journal of Multivariate Analysis, 100}, 518-532.
\item[] Srivastava, M. S., and Du, M. 2008. A test for the mean vector with fewer observations than the dimension. {\it Journal of Multivariate Analysis, 99}, 386-402.
\item[] Variyath, A. M., Vattathoor, J. 2014. Robust Control Charts for Monitoring Process Mean of Phase-I Multivariate Individual Observations. {\it Quality and Reliability Engineering International, 30}, 795-812.
\item[] Vargas, N. J. A. (2003). Robust estimation in multivariate control charts for individual observations. {\it Journal of Quality Technology, 35}, 367-376.
\item[] Walker, E., Wright, S. P. 2002. Comparing Curves Using Additive Models. {\it Journal of Quality Technology, 34}, 118–129.
\item[] Wang, K., Jiang, W. 2009. High-dimensional process monitoring and fault isolation via
variable selection. {\it Journal of Quality Technology, 41}, 247–258.
\item[] Welch, B. L. 1951. On the comparison of several mean values: an alternative approach. {\it Biometrika, 38}, 330-336.
\item[] Willems, G., Pison, G., Rousseeuw, P. J., Van Aelst, S. 2002. A robust Hotelling test. {\it Metrika, 55}, 125-138.
\item[] Williams, J. D., Woodall, W. H., \& Birch, J. B. 2003. Phase I monitoring of nonlinear profiles. {\it In quality and productivity research conference}, Yorktown Heights, New York.
\item[] Williams, J. D., Woodall, W. H., and Birch, J. B. 2007. Statistical monitoring of nonlinear product and process quality proﬁles. {\it Quality and Reliability Engineering International, 23}, 925–941.
\item[] Williams, J. D., Woodall, W. H., Birch, J. B., Sullivan, J. H. 2006. Distribution of Hotelling's $T^2$ statistic based on the successive differences estimator. {\it Journal of Quality Technology, 38}, 217-229.
\item[] Woodall, W. H. 2000. Controversies and contradictions in statistical process control (with discussion), {\it Journal of Quality Technology, 32}, 341–378.
\item[] Woodall, W. H., Montgomery, D. C. 2014. Some current directions in the theory and application of statistical process monitoring. {\it Journal of Quality Technology, 46}, 78-94.
\item[] Woodall, W. H., Spitzner, D. J., Montgomery, D. C., and Gupta, S. 2004. Using control charts to monitor
process and product quality proﬁles. {\it Journal of Quality Technology,  36}, 309–320.
\item[] Zhang, H., \& Albin, S. (2009). Detecting outliers in complex profiles using a $\chi^2$ control chart method. {\it IIE Transactions}, 41, 335-345.
\item[] Zhang, L., Zhu, T., Zhang, J. T. 2020. A simple scale-invariant two-sample test for high-dimensional data. {\it Econometrics and Statistics, 14}, 131-144.
\item[] Zou, C., Qiu, P. 2009. Multivariate statistical process control using LASSO. {\it Journal of
American Statistical Association, 104}, 1586–1596.
\end{enumerate}

\appendix
\newpage

\section*{Appendix A}
In this appendix, a consistent and unbiased estimator of $\textrm{tr}\,\bfmath{\rho}^{3}$ based on the $m$ samples available in Phase I will be presented. We first obtain expressions for $(\textrm{tr}\,\mathbf{S})^3$ and $\textrm{tr}\,\mathbf{S}^3$ in terms of chi-square random variables in a similar fashion as in Srivastava [2005], where expressions for $\textrm{tr}\,\mathbf{S}$, $\textrm{tr}\,\mathbf{S}^2$, and $(\textrm{tr}\,\mathbf{S})^2$ were derived, and their corresponding expectations will be then calculated.

From the spectral decomposition of the positive definite matrix $\mathbf{\Sigma}$, we can write $\mathbf{P}\mathbf{\Sigma}\mathbf{P}^\prime=\mathbf{\Xi}$, where $\mathbf{\Xi}$ is the diagonal matrix of the eigenvalues $\eta_{_1},\,\eta_{_2},\,\dots ,\,\eta_{_p}$ of $\mathbf{\Sigma}$ and  $\mathbf{P}$ is the  orthogonal matrix of the corresponding eigenvectors. Notice that  $\sum_{j=1}^{p}\eta_{_j}^k={\rm tr}(\mathbf{\Sigma}^k)$. To prove Theorem \ref{R3}, we follow the notations in Srivastava (2005). Recall that $\mathbf{X}_{1},\,\mathbf{X}_{2},\,\dots,\,\mathbf{X}_{m} \overset{i.i.d.}{\sim}N_p(\bfmath{\mu},\bfmath{\Sigma})$ and define the $p\times 1$ vectors  $\mathbf{Y}_i=\sqrt{m/(m-1)}\,(\mathbf{X}_{_i}-\overline{\mathbf{X}})$ and $\mathbf{U}_i=\mathbf{\Sigma}^{-1/2}\mathbf{Y}_i$ for $i=1,\,\dots,\,m$. Notice that $\mathbf{Y}_1,\,\dots,\,\mathbf{Y}_m \sim N_p(\mathbf{0},\,\bfmath{\Sigma})$ and $\mathbf{U}_1,\,\dots,\,\mathbf{U}_m \overset{i.i.d.}{\sim}N_p(\mathbf{0},\,\mathbf{I}_p)$. Define the $p\times m$ matrices $\mathbf{Y}=(\mathbf{Y}_1,\,\dots,\,\mathbf{Y}_m)$ and   $\mathbf{U}=(\mathbf{U}_1,\,\dots,\,\mathbf{U}_m)$ and notice that $\mathbf{Y}=\mathbf{\Sigma}^{1/2}\mathbf{U}$ and $m\,\mathbf{S}=\mathbf{Y}\mathbf{Y}'$. In addition, define the $m\times p$ matrix  $\mathbf{W}'=(\mathbf{W}_1,\,\dots,\,\mathbf{W}_p)=\mathbf{U}'\mathbf{P}'$, where $\mathbf{W}_1,\,\dots,\,\mathbf{W}_p \overset{i.i.d.}{\sim}N_m(\mathbf{0},\,\mathbf{I}_m)$.
Thus one can rewrite $\textrm{tr}\,\mathbf{S}$ as 
\begin{equation}\label{trSdecom}
{\rm tr}(\mathbf{S})=\frac{1}{m}{\rm tr}(\mathbf{Y}\,\mathbf{Y}')=\frac{1}{m}{\rm tr}(\mathbf{U}^\prime\mathbf{\Sigma}_{_0}\mathbf{U})=\frac{1}{m}{\rm tr}(\mathbf{W}'\mathbf{\Xi}\mathbf{W})=\frac{1}{m}\sum_{i=1}^{p}\eta_{_i}\mathbf{W}_{i}'\mathbf{W}_{i}\,.
\end{equation}
Define the random vraibales $V_{ij}=\mathbf{W}_{i}'\mathbf{W}_{j}$ and notice that $V_{ij}=V_{ji}$, $V_{ii}\,V_{ij}^2=V_{ji}\,V_{ii}\,V_{ij}$ and $V_{ii}$ are i.i.d. chi-square random variables with $m$ degrees of freedom. 
Since $\mathbf{W}_1,\,\dots,\,\mathbf{W}_p \overset{i.i.d.}{\sim}N_m(\mathbf{0},\,\mathbf{I}_m)$, we have the following moments that will be used in the remaining part of the proof. 
\begin{align}
{\rm E}\left[V_{ii}^r\right]&=m(m+2),\dots,(m+2r-2) \quad \textrm {for} \quad r=1,2,3,\dots\,,\label{chisq-moments}\\
    {\rm E}\left[V_{jk}^2\right]&={\rm tr}\left({\rm E}\left[\mathbf{W}_j\mathbf{W}_j'\right]\,{\rm E}\left[\mathbf{W}_k\,\mathbf{W}_k'\right]\right)={\rm tr}\left(\mathbb{I}_m\right)=m
    ={\rm E}\left[V_{jj}\right]\,,\label{EVjk2}\\
    {\rm E}\left[V_{ii}\,V_{jk}^2\right]&={\rm E}\left[V_{ii}\right]\,{\rm E}\left[V_{jk}^2\right]=m^2\,,\label{EViiVjk2}\\
    {\rm E}\left[V_{ii}\,V_{jj}^2\right]&={\rm E}\left[V_{ii}\right]{\rm E}\left[V_{jj}^2\right]=m^2(m+2)\,,\label{EVii2Vjj}\\
    {\rm E}\left[V_{ii}^2\,V_{ij}\right]&={\rm E}\left[V_{ii}^2\,\mathbf{W}_i'\right]\,{\rm E}\left[\mathbf{W}_j\right]=0\,,\label{EVii2Vij}\\
       {\rm E}\left[V_{ii}\,V_{ij}^2\right]&={\rm E}\left[V_{ii}\,V_{ij}\,V_{ji}\right]
={\rm E}\left[\mathbf{W}_i'\mathbf{W}_i\,\mathbf{W}_i'\mathbf{W}_j\,\mathbf{W}_j'\mathbf{W}_i\right]\label{EViiVij2}\\&={\rm E}\left[{\rm E}\left[\mathbf{W}_i'\mathbf{W}_i\,\mathbf{W}_i'\mathbf{W}_j\,\mathbf{W}_j'\mathbf{W}_i\mid \mathbf{W}_i\right]\right]
={\rm E}\left[\mathbf{W}_i'\mathbf{W}_i\,\mathbf{W}_i'\,{\rm E}\left[\mathbf{W}_j\,\mathbf{W}_j'\right]\mathbf{W}_i\right]\nonumber\\
&={\rm E}\left[\mathbf{W}_i'\mathbf{W}_i\,\mathbf{W}_i'\mathbf{W}_i\right]={\rm E}\left[V_{ii}^2\right]=m(m+2)\,,\nonumber\\
{\rm E}\left[V_{ii}\,V_{jj}\,V_{kk}\right]&={\rm E}\left[V_{ii}\right]{\rm E}\left[V_{jj}\right]{\rm E}\left[V_{kk}\right]=m^3\,,\label{EViiVjjVkk}\\
{\rm E}\left[V_{ki}\,V_{ij}\,V_{jk}\right]&={\rm E}\left[\mathbf{W}_k'\mathbf{W}_i\,\mathbf{W}_i'\mathbf{W}_j\,\mathbf{W}_j'\mathbf{W}_k\right]
={\rm E}\left[{\rm E}\left[\mathbf{W}_k'\mathbf{W}_i\,\mathbf{W}_i'\mathbf{W}_j\,\mathbf{W}_j'\mathbf{W}_k\mid \mathbf{W}_k\right]\right]\label{EVkiVijVjk}\\
&={\rm E}\left[\mathbf{W}_k'\,{\rm E}\left[\mathbf{W}_i\,\mathbf{W}_i'\right]{\rm E}\left[\mathbf{W}_j\,\mathbf{W}_j'\right]\mathbf{W}_k\right]\nonumber\\
&={\rm E}\left[\mathbf{W}_k'\,\mathbf{W}_k\right]={\rm E}\left[V_{kk}\right]=m\nonumber
\end{align}
In a similar fashion and using the results on the moments of quadratic forms and chi-square distribution (see Srivastava and Yanagihara, 2010), we can obtain the following higher order moments. Proofs are long and omitted. 
\begin{align}
{\rm Var}\left[V_{ii}^3\right]&=6\,m\,(m + 2)\,(m + 4)\,(3\,m^2 + 30\,m + 80)\,,\label{VarVii3}\\
{\rm E}\left[V_{ij}^4\right]&=3\,m\,(m+2)\,,\\
{\rm E}\left[V_{ii}^2\,V_{ij}^4\right]&=3\,{\rm E}\left[V_{ii}^4\right]=3\,m\,(m+2)\,(m+4)\,(m+6)\,,\\
{\rm E}\left[V_{ii}^4\,V_{jj}^2\right]&=m^2\,(m+2)^2\,(m+4)\,(m+6)\,,\\
{\rm E}\left[V_{ii}^3\,V_{jj}\,V_{ij}^2\right]&=3\,m\,(m + 2)\,(m + 4)(m + 6)\,,\\
{\rm E}\left[V_{ij}^2\,V_{jk}^2\,V_{ki}^2\right]&=m\,(m + 2)\,(m + 8)\,\\
{\rm E}\left[V_{ij}^3\,V_{jk}\,V_{ki}\,V_{kk}\right]&= 3\,m\,(m + 2)^2
\end{align}
%%%%%%%%%%%%%%%%%%%%%%%%
In addition, since
\begin{equation}\label{trSthen3}
\left(m\,{\rm tr}(\mathbf{S})\right)^3=\left(\sum_{i=1}^{p}\eta_{i}V_{ii}\right)^3=\sum_{i=1}^{p}\eta_{i}^3V_{ii}^3+3\sum_{i\neq {j}}\eta_{i}^2\eta_{j}V_{ii}^2V_{jj}+\sum_{i\neq {j}\neq {k}}\eta_{i}\eta_{j}\eta_{k}V_{ii}V_{jj}V_{kk}\,,
\end{equation}
from \eqref{chisq-moments}, \eqref{EVii2Vjj} and \eqref{EViiVjjVkk}, we have
\begin{equation}\label{EtrS3}
{\rm E}\left[\left(m\,{\rm tr}(\mathbf{S})\right)^3\right]=
\sum_{i=1}^{p}\eta_{i}^3m(m+2)(m+4)+3 \sum_{i\neq{j}}\eta_{i}^2\eta_{j}m^2(m+2)+\sum_{i\neq{j}\neq{k}}\eta_{i}\eta_{j}\eta_{k}m^3\,.
\end{equation}
To calculate  ${\rm tr}(\mathbf{S}^3)$ and ${\rm E}\left[{\rm tr}(\mathbf{S}^3)\right]$, we first write
\begin{align}\label{trS3}
m^3{\rm tr}\left(\mathbf{S}^3\right)=&{\rm tr}\left(\mathbf{W}'\mathbf{\Xi}\mathbf{W}\right)^3={\rm tr}\left(\sum_{i=1}^{p}\eta_i\,\mathbf{W}_i\mathbf{W}_i'\right)^3\nonumber\\
=&{\rm tr}\left(\sum_{i=1}^{p}\eta_{i}^3\left(\mathbf{W}_i\mathbf{W}_i'\right)^3\right)+3\,{\rm tr}\left(\sum_{i\neq j}\eta_{i}^2\eta_{j}\left(\mathbf{W}_i\mathbf{W}_i'\right)^2\left(\mathbf{W}_j\mathbf{W}_j'\right)\right)\nonumber\\
&+{\rm tr}\left(\sum_{i\neq j\neq k}\eta_{i}\eta_{j}\eta_{k}\left(\mathbf{W}_i\mathbf{W}_i'\right)\left(\mathbf{W}_j\mathbf{W}_j'\right)\left(\mathbf{W}_k\mathbf{W}_k'\right)\right)\nonumber \\
=&C_{1}+3\,C_2+C_3
\end{align}
and analyze each term $C_{1},\,C_2,\,C_3$ as follows:
\begin{align*}
C_{1}&={\rm tr}\left(\sum_{i=1}^{p}\eta_{i}^3\left(\mathbf{W}_i\mathbf{W}_i'\right)^3\right)={\rm tr}\left(\sum_{i=1}^{p}\eta_{i}^3\left(\mathbf{W}_i'\mathbf{W}_i\right)^3\right)
=\sum_{i=1}^{p}\eta_{i}^3\left(\mathbf{W}_i'\mathbf{W}_i\right)^3=\sum_{i=1}^{p}\eta_{i}^3V_{ii}^3\\
C_2&={\rm tr}\left(\sum_{i\neq j}\eta_{i}^2\eta_{j}\left(\mathbf{W}_i\mathbf{W}_i'\right)^2\left(\mathbf{W}_j\mathbf{W}_j'\right)\right)={\rm tr}\left(\sum_{i\neq j}\eta_{i}^2\eta_{j}\mathbf{W}_j'\left(\mathbf{W}_i\mathbf{W}_i'\right)^2\mathbf{W}_j\right)\\
&={\rm tr}\left(\sum_{i\neq j}\eta_{i}^2\eta_{j}\left(\mathbf{W}_j'\mathbf{W}_i\right)\left(\mathbf{W}_i'\mathbf{W}_i\right)\left(\mathbf{W}_i'\mathbf{W}_j\right)\right)=\sum_{i\neq j}\eta_{i}^2\eta_{j}V_{ii}\,V_{ij}^2\\
C_3&={\rm tr}\left(\sum_{i\neq j\neq k}\eta_{i}\eta_{j}\eta_{k}\left(\mathbf{W}_i\mathbf{W}_i'\right)\left(\mathbf{W}_j\mathbf{W}_j'\right)\left(\mathbf{W}_k\mathbf{W}_k'\right)\right)=\sum_{i\neq j\neq k}\eta_{i}\eta_{j}\eta_{k}V_{ki}\,V_{ij}\,V_{jk}\,.
\end{align*}
Hence using equations  \eqref{chisq-moments}, \eqref{EViiVij2} and \eqref{EVkiVijVjk}, 
\begin{equation}\label{ES3}
{\rm E}\left[m^3(\textrm{tr}\,\mathbf{S}^3)\right]
=m(m+2)(m+4)\sum_{i=1}^{p}\eta_{i}^3+3\,m(m+2)\sum_{i\neq j}^p\eta_{i}^2\eta_{j}+m\sum_{i\neq j\neq k}^p\eta_{i}\eta_{j}\eta_{k}\,.
\end{equation}
Equations \eqref{EtrS3} and \eqref{ES3} lead to the following identity    
%To derive an estimator for $\textrm{tr}\,\mathbf{\Sigma}_{_0}^3/p$, we notice that
\begin{align*}
{\rm E}\left[{\rm tr}\left(\mathbf{S}^3\right)-\frac{1}{m^2}\left({\rm tr}\left(\mathbf{S}\right)\right)^3\right]&=
\frac{(m+2)(m-1)}{m^3}\left[\frac{(m+4)(m+1)}{m}\sum_{i=1}^{p}\eta_{i}^3+3\sum_{i\neq{j}}\eta_{i}^2\eta_{j}\right]\\
&=\frac{(m+2)(m-1)}{m^3}\left[\frac{(m+4)(m+1)}{m}\sum_{i=1}^{p}\eta_{i}^3+3\left(\sum_{i=1}^{p}\eta_{i}^2\sum_{i=1}^{p}\eta_{i}-\sum_{i=1}^{p}\eta_{i}^3\right)\right]\\
&=\frac{(m+2)(m-1)}{m^3}\,\left[\frac{(m^2+2m+4)}{m}\,{\rm tr}\left({\bfmath{\Sigma}^3}\right)+3\,{\rm tr}\left(\bfmath{\Sigma}\right)\,{\rm tr}\left(\bfmath{\Sigma}^2\right)\right]\,.
\end{align*}
This identity, in turn, suggests a way to obtain an unbiased estimator of ${\rm tr}\left(\mathbf{\Sigma}^3\right)$. Srivastava (2005) proposed unbiased estimators of ${\rm tr}\left(\mathbf{\Sigma}\right)$ and ${\rm tr}\left(\mathbf{\Sigma}^2\right)$ as 
\begin{equation*}
{\rm tr}\left(\mathbf{S}\right) \quad \text{ and } \quad 
\frac{m^2}{(m-1)(m+2)}\left[{\rm tr}\left(\mathbf{S}^2\right)-m^{-1}\left({\rm tr}\left(\mathbf{S}\right)\right)^2\right]\,,
\end{equation*}
respectively, as well as showed that 
\begin{equation*}
{\rm Cov}\left({\rm tr}\left(\mathbf{S}\right),\,{\rm tr}\left(\mathbf{S}^2\right)-m^{-1}\,{\rm tr}\left(\mathbf{S}\right)^2\,\right)=\frac{4(m-1)(m+2)}{m^3}{\rm tr}\left(\bfmath{\Sigma}^3\right)\,.
\end{equation*}
So, with a little bit of algebra, we have 
\begin{align*}
{\rm E}\left[ {\rm tr}\left(\mathbf{S}\right)\cdot\left({\rm tr}\left(\mathbf{S}^2\right)-m^{-1}\left({\rm tr}\left(\mathbf{S}\right)\right)^2 \right)\right]=\frac{(m-1)(m+2)}{m^2}\left[\frac{4}{m}\, {\rm tr}\left(\bfmath{\Sigma}^3 \right)+{\rm tr}\left(\bfmath{\Sigma} \right)\,{\rm tr}\left(\bfmath{\Sigma}^2 \right) \right]\,.
\end{align*}
Then, an unbiased estimator for $p^{-1}{\rm tr}\left({\bfmath{\Sigma}^3}\right)$ is
\begin{equation}\label{EstSig3}
\frac{m^4}{(m+4)(m-1)(m^2-4)}\cdot \frac{1}{p}\left[{\rm tr}\left(\mathbf{S}^3\right)-\frac{3}{m}\,{\rm tr}\left(\mathbf{S}\right){\rm tr}\left(\mathbf{S}^2\right)+\frac{2}{m^2}\,\left({\rm tr}\left(\mathbf{S}\right)\right)^3\right]\,.
\end{equation} 
In the remaining part, we show that the estimator in \eqref{EstSig3}, or equivalently
\begin{equation}\label{consist_J}
J=\frac{1}{p}\left[{\rm tr}\left(\mathbf{S}^3\right)-\frac{3}{m}\,{\rm tr}\left(\mathbf{S}\right){\rm tr}\left(\mathbf{S}^2\right)+\frac{2}{m^2}\,\left({\rm tr}\left(\mathbf{S}\right)\right)^3\right]
\end{equation} 
is a consistent estimator for $p^{-1}{\rm tr}\left({\bfmath{\Sigma}^3}\right)$ when $m\rightarrow\infty$. For this purpose, it suffices to show that the expectation of $J$ goes to $p^{-1}{\rm tr}\left({\bfmath{\Sigma}^3}\right)$, while its variance goes to zero when $m\rightarrow\infty$. Recall equations \eqref{trSdecom} and \eqref{trS3}. In a similar way, we can show that
\begin{align}\label{trS2}
m^2{\rm tr}\left(\mathbf{S}^2\right)&={\rm tr}\left(\mathbf{W}'\mathbf{\Xi}\mathbf{W}\right)^2 \nonumber\\
&={\rm tr}\left(\sum_{i=1}^{p}\eta_{i}^2\left(\mathbf{W}_i\mathbf{W}_i'\right)^2\right)+{\rm tr}\left(\sum_{i\neq j}\eta_{i}\eta_{j}\left(\mathbf{W}_i\mathbf{W}_i'\right)\left(\mathbf{W}_j\mathbf{W}_j'\right)\right)\\
&=\sum_{i=1}^{p}\eta_{i}^2V_{ii}^2+\sum_{i\neq j}\eta_{i}\eta_{j}V_{ij}^2\,,\nonumber
\end{align}
and 
\begin{align}\label{trStrS2}
m^3\,{\rm tr}\left(\mathbf{S}\right){\rm tr}\left(\mathbf{S}^2\right)&=
\left(\sum_{i=1}^{p}\eta_{_i}V_{ii}\right)\left(\sum_{i=1}^{p}\eta_{_i}^2V_{ii}^2+\sum_{i\neq j}\eta_{_i}\eta_{_j}V_{ij}^2\right)\nonumber\\
&=\sum_{i=1}^{p}\eta_{_i}^3V_{ii}^3+\sum_{i\neq j}\eta_{_i}\eta_{_j}\left(\eta_{_j}V_{ii}V_{jj}^2+2\,\eta_{_i}V_{ii}V_{ij}^2 \right)+\sum_{i\neq j \neq k}\eta_{_i}\eta_{_j}\eta_{_k}V_{ii}V_{jk}^2
\end{align}
Using \eqref{trSthen3}, \eqref{trS3} and \eqref{trStrS2}, the estimator $J$ in \eqref{consist_J} can be written as
\begin{align*}
J=\frac{1}{m^3p}&\left[\frac{(m-2)(m-1)}{m^2}\sum_{i=1}^{p}\eta_{_i}^3V_{ii}^3+\frac{3(m-2)}{m}\sum_{i\neq j}\eta_{_i}^2\eta_{_j}\left(V_{ii}V_{ij}^2-\frac{1}{m}V_{ii}^2V_{jj}\right)\right.\\
&\left.\sum_{i\neq j \neq k}\eta_{_i}\eta_{_j}\eta_{_k}\left(V_{ki}V_{ij}V_{jk}-\frac{3}{m}V_{ii}V_{jk}^2+\frac{2}{m^2}V_{ii}V_{jj}V_{kk}\right) \right]
\end{align*}
From this representation and equations \eqref{chisq-moments}, \eqref{EViiVjk2},
\eqref{EVii2Vjj},
\eqref{EViiVij2}, 
\eqref{EViiVjjVkk},
\eqref{EVkiVijVjk},   we obtain
\begin{align*}
{\rm E}\left[ J\right]&=\frac{1}{m^3p}\left[\frac{(m-2)(m-1)}{m^2}\sum_{i=1}^{p}\eta_{_i}^3{\rm E}\left[V_{ii}^3\right]+\frac{3(m-2)}{m}\sum_{i\neq j}\eta_{_i}^2\eta_{_j}\left({\rm E}\left[ V_{ii}V_{ij}^2\right]-\frac{1}{m}{\rm E}\left[V_{ii}^2V_{jj}\right]\right)\right.\\
&\qquad \qquad \left.\sum_{i\neq j \neq k}\eta_{_i}\eta_{_j}\eta_{_k}\left({\rm E}\left[V_{ki}V_{ij}V_{jk}\right]-\frac{3}{m}{\rm E}\left[V_{ii}V_{jk}^2\right]+\frac{2}{m^2}{\rm E}\left[V_{ii}V_{jj}V_{kk}\right]\right) \right]\\
&=\frac{(m+4)(m^2-4)(m-1)}{m^4}\cdot \frac{1}{p}\,\sum_{i=1}^{p}\eta_{_i}^3\,.
\end{align*}
Obviously ${\rm E}\left[J\right]\rightarrow p^{-1}\sum_{i=1}^{p}\eta_{_i}^3=p^{-1}{\rm tr}\left(\mathbf{\Sigma}^3\right)$ as $m\rightarrow \infty$. 

\noindent 

Now, to compute the variance of $J$, we first define
\begin{align*}
J_1&=\sum_{i=1}^{p}\eta_{_i}^3V_{ii}^3\,,\\
J_2&=\sum_{i\neq j}\eta_{_i}^2\eta_{_j}\left(V_{ii}V_{ij}^2-\frac{1}{m}V_{ii}^2V_{jj}\right)\,,\\
J_3&=\sum_{i\neq j \neq k}\eta_{_i}\eta_{_j}\eta_{_k}\left(V_{ki}V_{ij}V_{jk}-\frac{3}{m}V_{ii}V_{jk}^2+\frac{2}{m^2}V_{ii}V_{jj}V_{kk}\right)
\end{align*}
and notice that
\begin{equation}\label{Jin123}
J=\frac{1}{m^3p}\left[\frac{(m-2)(m-1)}{m^2}\,J_1+\frac{3(m-2)}{m}\,J_2
+J_3 \right]
\end{equation}
So, the variance of $J$ can be written as
\begin{align*}
{\rm Var}\left[J \right]&=\frac{1}{m^6p^2}\left[\frac{(m-2)^2(m-1)^2}{m^4}\,{\rm Var}\left[J_1\right]+\frac{9(m-2)^2}{m^2}\,{\rm Var}\left[J_2\right]+{\rm Var}\left[J_3\right]\right.\\
&\qquad \qquad\left.+\frac{6(m-2)^2(m-1)}{m^3}\, \,{\rm Cov}\left[ J_1,\,J_2\right]+\frac{2(m-2)(m-1)}{m^2}\,{\rm Cov}\left[J_1,\,J_3 \right]\right.\\
&\qquad \qquad \left.+\frac{6(m-2)}{m}\,{\rm Cov}\left[J_2,\,J_3 \right]\right]
\end{align*}
We now compute the variance and covariances of $J_1$, $J_2$ and $J_3$. 
It is easy to see that
\begin{align*}
{\rm Var}\left[J_1\right]=\sum_{i=1}^{p}\eta_{_i}^6\,{\rm Var}\left[V_{ii}^3\right]&=6\,m(m+2)(m+4)(3\,m^2+30\,m+80)\,\sum_{i=1}^{p}\eta_{_i}^6\\
&=O(m^5)\sum_{i=1}^{p}\eta_{_i}^6=O(m^5)\,{\rm tr}\left(\mathbf{\Sigma}^6 \right)\,.
\end{align*}
Set $K_{ij}=V_{ii}V_{ij}^2-m^{-1}\,V_{ii}^2V_{jj}$ and note that from equations \eqref{chisq-moments}, \eqref{EVii2Vjj}, \eqref{EViiVij2} for $i\neq j$, $r\neq \ell$ and $(i,\,j)\neq (r,\,\ell)$, we have ${\rm E}\left[K_{ij} \right]=0$, ${\rm Cov}\left[K_{ij},\,K_{r \ell}\right]=0$, and 
${\rm Var}\left[K_{ij}\right]=m^{-1}(m-2)\,(m-1)\,{\rm E}\left[V_{ii}^4 \right]$. Thus from \eqref{chisq-moments}
\begin{align*}
{\rm Var}\left[J_2\right]=\sum_{i\neq j}^p\eta_{_i}^4\,\eta_{_j}^2\,{\rm Var}\left[K_{ij}\right]&=\sum_{i\neq j}^p\eta_{_i}^4\,\eta_{_j}^2\,m^{-1}(m-2)\,(m-1)\,{\rm E}\left[V_{ii}^4 \right]\\
&=O(m^5) \,\sum_{i\neq j}^p\eta_{_i}^4\,\eta_{_j}^2 =O(m^5) \,\left[{\rm tr}\left(\mathbf{\Sigma}^4 \right){\rm tr}\left(\mathbf{\Sigma}^2 \right)-{\rm tr}\left(\mathbf{\Sigma}^6 \right)\right]\,.
\end{align*}
Finally, define $S_{ijk}=V_{ij}V_{jk}V_{ki}-3\,m^{-1}\,V_{ii}V_{jk}^2+2\,m^{-2}\,V_{ii}\,V_{jj}\,V_{kk}$, so from equations \eqref{EViiVjk2}, \eqref{EViiVjjVkk}, \eqref{EVkiVijVjk}  for $i,\, j,\, k,\, r,\,\ell,\,q$ all different and $(i,\,j,\,k)\neq (r,\,\ell,\,q)$ we have ${\rm E}\left[S_{ijk} \right]=0$, ${\rm Cov}\left[S_{ijk},\,K_{r \ell q}\right]=0$ and 
${\rm Var}\left[S_{ijk}\right]=O(m^3)$. Thus 
\begin{align*}
{\rm Var}\left[J_3\right]&=\sum_{i\neq j \neq k}\eta_{_i}^2\eta_{_j}^2\eta_{_k}^2 {\rm Var}\left[S_{ijk} \right]\\
&=O(m^3) \,\sum_{i\neq j \neq k}\eta_{_i}^2\eta_{_j}^2\eta_{_k}^2\\
&=O(m^3) \,\left[\left({\rm tr}\left(\mathbf{\Sigma}^2 \right)\right)^3+2\,{\rm tr}\left(\mathbf{\Sigma}^6 \right)-3\,{\rm tr}\left(\mathbf{\Sigma}^4 \right){\rm tr}\left(\mathbf{\Sigma}^2 \right)\right]\,.
\end{align*}
Additionally, we can show that all pairwise covariances of $J_1$, $J_2$ and $J_3$ are zero. Hence
\begin{align*}
{\rm Var}\left[J \right]=&O(m^{-1}p^{-1})\left(\frac{1}{p}\,{\rm tr}\left(\mathbf{\Sigma}^6 \right) \right)+
O(m^{-1})\left[\left(\frac{1}{p}\,{\rm tr}\left(\mathbf{\Sigma}^4 \right) \right)\left(\frac{1}{p}\,{\rm tr}\left(\mathbf{\Sigma}^2 \right) \right)-\left(\frac{1}{p^2}\,{\rm tr}\left(\mathbf{\Sigma}^6 \right) \right)\right]\\
&+O(p\,m^{-3})\,\left[\left(\frac{1}{p}\,{\rm tr}\left(\mathbf{\Sigma}^2 \right)\right)^3+\frac{2}{p^2}\,\left(\frac{1}{p}\,{\rm tr}\left(\mathbf{\Sigma}^6 \right)\right)-\frac{3}{p}\,\left(\frac{1}{p}\,{\rm tr}\left(\mathbf{\Sigma}^4 \right)\right)\,\left(\frac{1}{p}\,{\rm tr}\left(\mathbf{\Sigma}^2 \right)\right)\right]\,.
\end{align*}
This result proves that when $m=O(p^\delta)$ for $\delta>1/3$, ${\rm Var}\left[J \right]\rightarrow 0$, as $m,\,p\rightarrow \infty$. Thus we proved consistency of the estimator $J$ for $p^{-1}{\rm tr}\left(\mathbf{\Sigma}^3\right)$.

\noindent
Now, we show that replacing $\mathbf{S}$ with $\mathbf{R}$ in estimator \eqref{consist_J} can give us a consistent estimator for $p^{-1}{\rm tr}\left(\bfmath{\rho}^{3}\right)$ of the form
\begin{align}\label{est.rho3}
\frac{1}{p}\,\left[{\rm tr}\left(\mathbf{R}^3\right)-\frac{3\,p}{m}\,{\rm tr}\left({\mathbf{R}^2}\right)+\frac{2\,p^3}{m^2}\right]\,.
\end{align}
We first notice that if the observations are  $\mathbf{D}^{-1/2}X_1,\,\dots,\,\mathbf{D}^{-1/2}X_m\overset{i.i.d.}{\sim}N_p(\bfmath{\mu},\bfmath{\rho})$, rather than $\mathbf{X}_{1},\,\mathbf{X}_{2},\,\dots,\,\mathbf{X}_{m} \overset{i.i.d.}{\sim}N_p(\bfmath{\mu},\bfmath{\Sigma})$, then the estimator in \eqref{consist_J} is consistent for $p^{-1}{\rm tr}\left(\bfmath{\rho}^{3}\right)$. So we can simply assume that $\mathbf{\Sigma}=\bfmath{\rho}$. We also notice that the sample correlation matrix $\mathbf{R}$ is invariant under scale transformation and so is the estimator \eqref{est.rho3}. Motivated by Srivastava and Du (2008), let $\omega_i=1-1/s_{ii}$, then $\omega_i=O_p(n^{-\frac{1}{2}})$, and $\mathbf{D}_{_{S}}^{-1}=\mathbf{I}-\mathbf{D}_{\mathbf{\omega}}$, 
where $\mathbf{D}_{\mathbf{\omega}}={\rm diag}(\omega_1,\,\dots,\,\omega_p)$. Then

{\footnotesize\begin{align*}
\frac{1}{p}\left[{\rm tr}\left(\mathbf{R}^3\right)-\frac{3p}{m}{\rm tr}\left({\mathbf{R}^2}\right)+\frac{2\,p^3}{m^2}\right]=&
\frac{1}{p}\,\left[{\rm tr}\left(\mathbf{D}_{_{S}}^{-1}\mathbf{S}\right)^3-\frac{3\,{\rm tr}\left(\mathbf{D}_{_{S}}^{-1}\mathbf{S}\right)}{m}\,{\rm tr}\left(\mathbf{D}_{_{S}}^{-1}\mathbf{S}\right)^2+2\,\frac{({\rm tr}\left(\mathbf{D}_{_{S}}^{-1}{\mathbf{S}}\right)^3}{m^2}\right]\\
=&\frac{1}{p}\left[{\rm tr}\left(\mathbf{S}^3\right)-3\,{\rm tr}\left(\mathbf{D}_{\mathbf{\omega}}\mathbf{S}^3\right)+
3\,{\rm tr}\left(\mathbf{D}_{\mathbf{\omega}}^2\mathbf{S}^3\right)-{\rm tr}\left(\mathbf{D}_{\mathbf{\omega}}^3\mathbf{S}^3\right)\right.\\
&-\frac{3\,{\rm tr}\left(\mathbf{S}-\mathbf{D}_{\mathbf{\omega}}\mathbf{S}\right)}{m}\left({\rm tr}\left(\mathbf{S}^2\right)-2{\rm tr}\left(\mathbf{D}_{\mathbf{\omega}}\mathbf{S}^2\right)+{\rm tr}\left(\mathbf{D}_{\mathbf{\omega}}\mathbf{S}\right)^2\right)\\
&\left.+\frac{2}{m^2}\left(\left({\rm tr}\left(\mathbf{S}\right)\right)^3+
3\,{\rm tr}\left(\mathbf{S}\right)\,\left({\rm tr}\left(\mathbf{D}_{\mathbf{\omega}}\mathbf{S}\right)\right)^2-3\,\left({\rm tr}\left(\mathbf{S}\right)\right)^2\,\left({\rm tr}\left(\mathbf{D}_{\mathbf{\omega}}\mathbf{S}\right)\right)-\left({\rm tr}\left(\mathbf{D}_{\mathbf{\omega}}\mathbf{S}\right)\right)^3\right)\right]
\\
=&\frac{1}{p}\left[{\rm tr}\left(\mathbf{S}^3\right)-\frac{3\,{\rm tr}\left(\mathbf{S}\right)}{m}\,{\rm tr}\left(\mathbf{S}^2\right)+2\,\frac{\left({\rm tr}\left(\mathbf{S}\right)\right)^3}{m^2}\right]\\
&+\frac{1}{p}\left[-3\,{\rm tr}\left(\mathbf{D}_{\mathbf{\omega}}\mathbf{S}^3\right)
+\frac{6}{m}\,{\rm tr}\left(\,\mathbf{S}\right)\,{\rm tr}\left(\mathbf{D}_{\mathbf{\omega}}\mathbf{S}^2\right)\right.\\
&\left.+\frac{3}{m}\,{\rm tr}\left(\mathbf{D}_{\mathbf{\omega}}\mathbf{S}\right)\,{\rm tr}\left(\mathbf{S}^2\right)-\frac{6}{m^2}\,\left({\rm tr}\left(\mathbf{S}\right)\right)^2\,\left({\rm tr}\left(\mathbf{D}_{\mathbf{\omega}}\mathbf{S}\right)\right)\right]\\
&+\frac{1}{p}\left[3\,{\rm tr}\left(\mathbf{D}_{\mathbf{\omega}}^2\mathbf{S}^3\right)-\frac{3}{m}\,{\rm tr}\left(\mathbf{S}\right){\rm tr}\left(\mathbf{D}_{\mathbf{\omega}}\mathbf{S}\right)^2\right.\\
&\left.-\frac{6}{m}\,{\rm tr}\left(\mathbf{D}_{\mathbf{\omega}}\mathbf{S}\right)\,{\rm tr}\left(\mathbf{D}_{\mathbf{\omega}}\mathbf{S}^2\right)+\frac{6}{m^2}\,{\rm tr}\left(\mathbf{S}\right)\,\left({\rm tr}\left(\mathbf{D}_{\mathbf{\omega}}\mathbf{S}\right)\right)^2\right]\\
&+\frac{1}{p}\,\left[-{\rm tr}\left(\mathbf{D}_{\mathbf{\omega}}^3\mathbf{S}^3\right)+\frac{3}{m}\,{\rm tr}\left(\mathbf{D}_{\mathbf{\omega}}\mathbf{S}\right)\,{\rm tr}\left(\mathbf{D}_{\mathbf{\omega}}\mathbf{S}\right)^2-\frac{2}{m^2}\,\left({\rm tr}\left(\mathbf{D}_{\mathbf{\omega}}\mathbf{S}\right)\right)^3\right]\,.
\end{align*}}
Under the condition (A.1) and when $m = O(p^{\zeta}), 0<\zeta\leq{1}$, from Theorem \ref{R3}, we have 
\begin{align*}
\frac{1}{p}[\textrm{tr}\,\mathbf{S}^3-\frac{3\textrm{tr}\,\mathbf{S}}{m}\textrm{tr}\,\mathbf{S}^2+2\frac{(\textrm{tr}\,\mathbf{\mathbf{S}})^3}{m^2}]-\frac{\textrm{tr}\bfmath{\rho}^{3}}{p}\overset{p}{\rightarrow}{0}, \quad (m, p)\rightarrow\infty
\end{align*}
the remaining terms in the decomposition above all go to zero as $m,\,p\rightarrow\infty$ while $m = O(p^{\zeta}), 0<\zeta\leq{1}$. This completes the proof of Theorem \ref{R3}.

\newpage
\section*{Appendix B}
In this Appendix, through a simulation study, we discuss the details on how we arrived in the formula for the finite sample correction coefficient \eqref{cpm}. Motivated by \eqref{uhat}, if we replace  ${\rm tr}(\bfmath{\rho}^2)$ by its estimate introduced in Section 2.1., our simulation study suggests fitting a simple linear function of $k$ in form of
\begin{equation}\label{cpmk}
c_{p,\,m}=1+\frac{p\cdot k}{m\,\sqrt{2\left[\rm tr\left(\bfmath{R}^{2}\right)-\frac{p^2}{m}\right]}},
\end{equation}
This type of simulation approach to finding a finite sample correction for test statistics has been used in literature. For example Pison et al. 2002 and Willems et al. 2002) investigated finite sample correction factors for affine invariant test statistics which suffices to generate samples from a standard multivariate normal distribution. In our setting, though, the statistic $U_{i}$ is invariant under translation and scale transformations. Additionally, since we are working under the sparsity of the covariance matrix ({Assumption  1}), we devise our simulation study in the following manner.  
\begin{itemize}
\item[{\bf 1.} ] Choose arbitrary values for $p$ and $m$, consider a multivariate normal process with known parameters $\bfmath{\mu}$ and $\mathbf{D}$, and generate $m$ multivariate normal observations based on these parameters.

\item[{\bf 2.} ] Compute the sample mean and sample diagonal matrix based on the generated sample in step 1. 

\item[{\bf 3.} ] Generate a large number of observations (e.g. $L=10000$) from the same distribution and compute the corresponding values of statistic $U_i$ based on the estimates in step 2. Sort these $U_i$ values and denote it by vector ${\bf A}_1$

\item[{\bf 4.} ] Generate $L$ observations from the univariate standard normal distribution. Modify each observation by adding the corresponding Cornish-Fisher value \eqref{CFexp1} based on the exact values of $\bfmath{\mu}$ and $\mathbf{D}$. Sort the obtained set and denote it by vector ${\bf A}_2$.

\item[{\bf 5.} ] Compute ${\bf A}={\bf A}_2/{\bf A}_1$ by element-wise division of the two vectors ${\bf A}_2$ and ${\bf A}_1$. Consider the average upper quantile (e.g $5\%$) of ${\bf A}$ as $c_{p,\,m}$.

\item[{\bf 5.} ] Fit a regression line \eqref{cpmk} between
\begin{equation}\label{reg}
y=c_{p,\,m}-1\,, \quad {\rm and} \quad x=\frac{p}{m\sqrt{2\left[\rm {tr}(\bfmath{R}^{2})-\frac{p^2}{m}\right]}}\,,
\end{equation}
where $\bfmath{R}$ is the sample correlation matrix calculated based on data generated data in step 1.
\end{itemize}
\begin{figure}[h!]
  \centering
  \begin{subfigure}[b]{0.6\linewidth}
    \includegraphics[width=\linewidth]{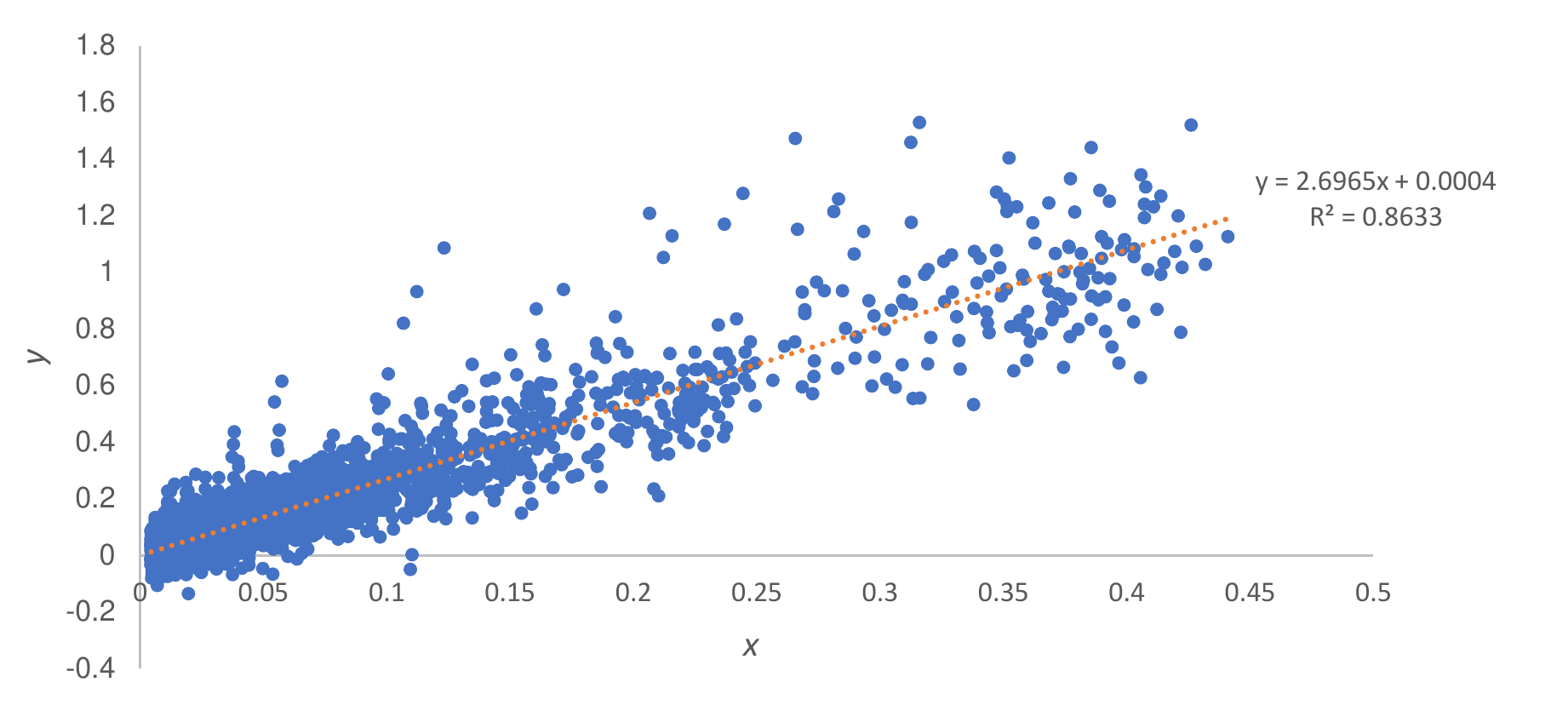}
    \caption{Scenario I}
  \end{subfigure}
  \begin{subfigure}[b]{0.6\linewidth}
    \includegraphics[width=\linewidth]{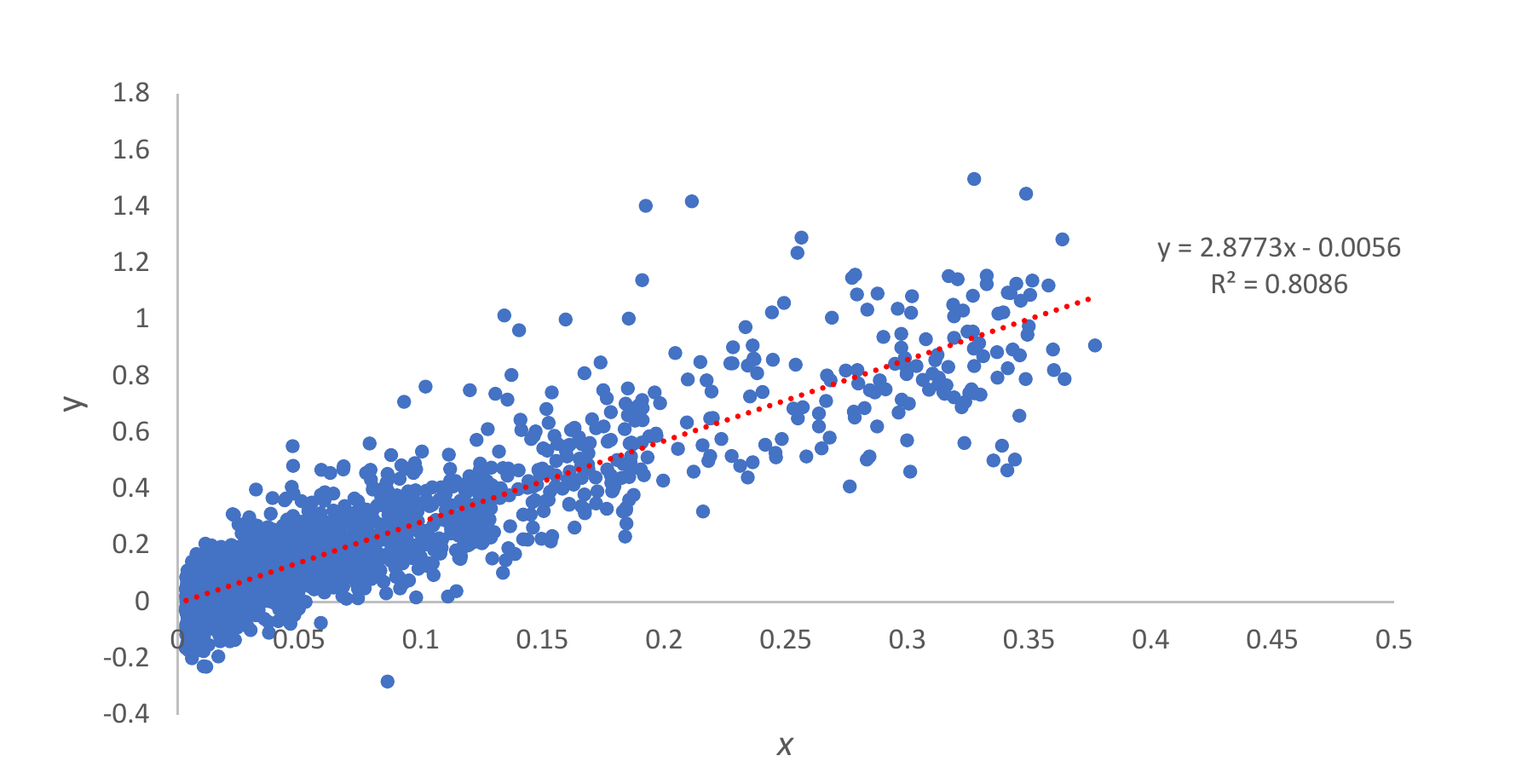}
    \caption{Scenario II}
  \end{subfigure}
    \begin{subfigure}[b]{0.6\linewidth}
    \includegraphics[width=\linewidth]{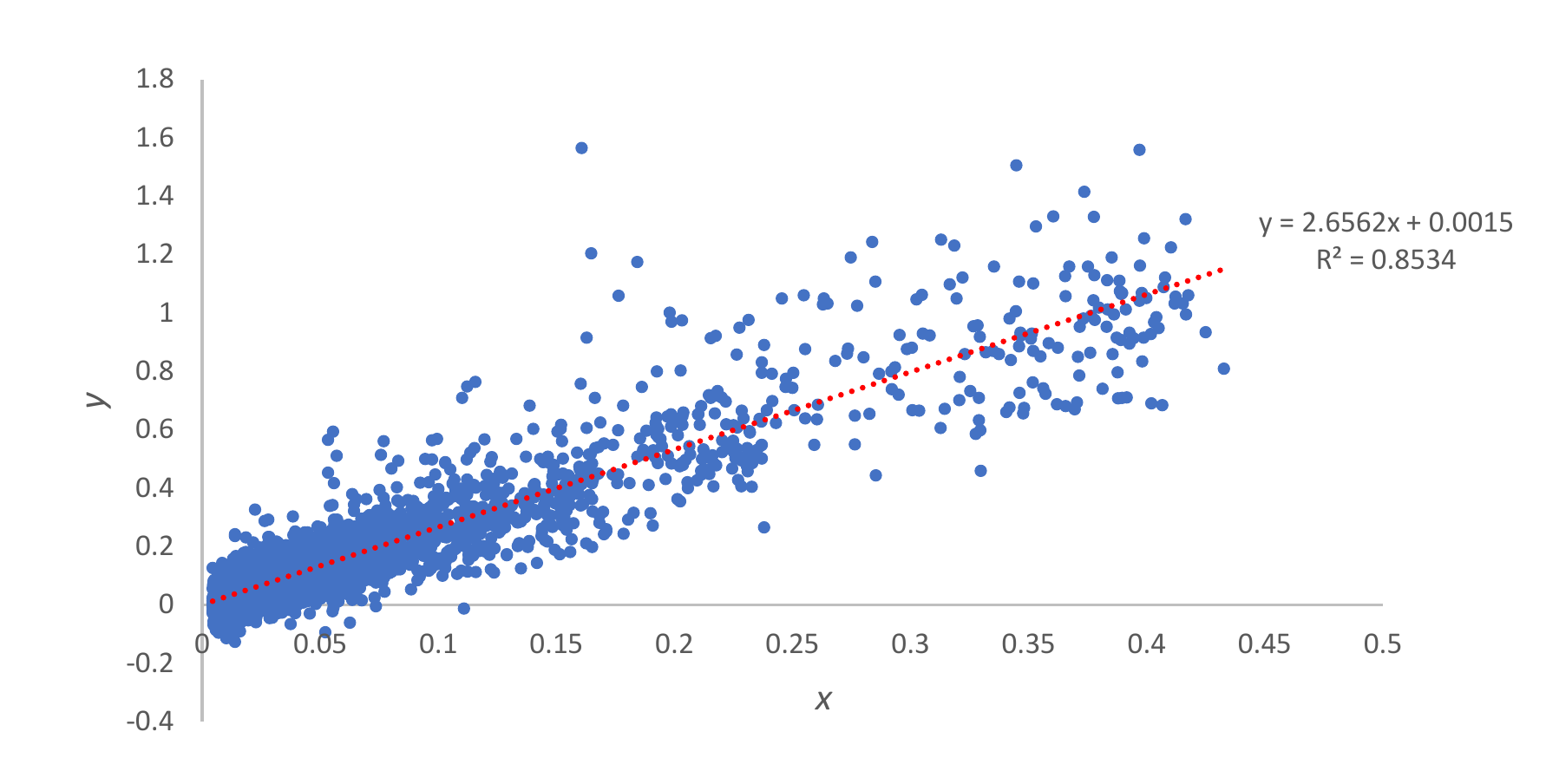}
    \caption{Scenario III}
  \end{subfigure}
  \caption{The fitted regression lines between $x$ and $y$ defined in \eqref{reg} for three multivariate normal distributions with different covariance matrices.}
  \label{fig:reg}
\end{figure}

Figure \ref{fig:reg} shows the fitted regression function between $x$ and $y$ for three multivariate normal distribution with different covariance matrix named as Scenario I, II, and III. For a p-variate process, all scenarios have the mean vector of $\mathbf{0}_{p}$. The covariance matrix of the first scenario is $\mathbf{I}_p$, for the second scenario it has the structure of an autoregressive correlation with $\sigma_{ij}=(0.5)^{|i-j|}\quad {\rm for}\quad i,j=1,\dots,p$, and in the third scenario the covariance matrix has elements $\sigma_{ii}={\rm Unif}(2,\,3)$ and $\sigma_{ij}={\rm exp}(-i-j),\quad {\rm for}\quad i\neq{j}$.
For the three scenarios described above, we generated 5000 values for $x$ and $y$ based on different pairs of sample size $m$ and dimensions $p$ and then a regression line is fitted between $x$ and $y$. For this purpose, $m$ and $p$ are considered to vary between 20-400 and 10-200, respectively.  In all cases, it can be concluded that the intercept is almost zero, slope of all regression lines are close to each other, and $R^2$ of regression lines are large enough. We evaluated other processes with different covariance structure and obtained a similar conclusion. Even for the cases that the covariance matrix is not sparse, a similar pattern for the regression line can be observed but $R^2$ might be smaller. For example, Table 1 shows the parameters of regression line when different magnitude of corrletion between variables is considered in the second scenrio's covariance sturcture. In other words, this table shows the regression coefficients for covariance matrix with $\sigma_{ij}=a^{|i-j|}\quad {\rm for}\quad i,j=1,\dots,p$ when $a$ varies between 0.1 and 0.9. It can be concluded from this table that, except when $a$ is very large, the values of $k$ are close for different values of $a$ while the intercepts of regression lines are almost zero. So, we propose using $k\approx{2.8}$ which leads us to the equation \eqref{e11} given in section 2.1. Our simulation showed that the chart performance is not sensitive to the misspecification of $k$. We show in Section 3 that using this function gives us the desired in-control probability of signal.

\begin{table}[h!]

\tiny
\centering
\caption{Parameters of regression line for cavariance matrix $\sigma_{ij}=(a)^{|i-j|}\quad {\rm for}\quad i,j=1,\dots,p$ with different values of $a$}
\label{tab1}
\begin{tabular}{ |p{1.1cm}|p{0.85cm}|p{0.85cm}|p{0.85cm}|p{0.85cm}|p{0.85cm}|p{0.85cm}|p{0.85cm}|p{0.85cm}|p{0.85cm}|  }
 \hline
{\centering{$a$}}&  {\centering{0.1}} & {\centering{0.2}}& {0.3} & {0.4}& {0.5} & {0.6}& {0.7} & {0.8}& {0.9}\\\hline
Intercept& 0.002&0.0026 &0.0051&-0.0075&-0.0056& -0.0097&-0.0192 &-0.0281&-0.0439\\ \hline 
Slope ($k$)&2.71 &2.76 &2.79 &2.82&2.88 &3.04&3.22 &3.70 &4.60\\ \hline 
$R^2$&0.87&0.84  &0.86 &0.87&0.81 &0.76&0.77 &0.6 &0.53\\ \hline 
\end{tabular}
\end{table}
%%%%%%%%%%%%%%%%%%%%%
\end {document}